\documentclass[a4paper,11pt]{article}
\pdfoutput=1 

\usepackage{jheppub} 

\usepackage[T1]{fontenc} 
\usepackage{cancel}

\title{\boldmath Entropy Current and Fluid-Gravity Duality in Gauss-Bonnet theory}

\author[a]{Chandranathan A,}
\author[b]{Sayantani Bhattacharyya,}
\author[b]{Milan Patra,}
\author[b]{and Shuvayu Roy}

\affiliation[a]{
International Centre for Theoretical Sciences (ICTS-TIFR),\\
Tata Institute of Fundamental Research, Shivakote, Hesaraghatta, Bangalore 560089, India.}
\affiliation[b]{School of Physical Sciences, National Institute of Science Education and Research\\
An OCC of Homi Bhabha National Institute, Jatani-752050, India}

\emailAdd{chandranathan.a@icts.res.in}
\emailAdd{sayanta@niser.ac.in}
\emailAdd{milan.patra@niser.ac.in}
\emailAdd{shuvayu.roy@niser.ac.in}

\abstract{Working within the approximation of small amplitude expansion, recently an entropy current has been constructed on the horizons of dynamical black hole solution in any higher derivative theory of gravity. In this note, we have dualized this horizon entropy current to a boundary entropy current in an asymptotically AdS black hole metric with a dual description in terms of dynamical fluids living on the AdS boundary.  This boundary entropy current is constructed using a set of mapping functions relating each point on the horizon to a point on the boundary. We have applied our construction to black holes in Einstein-Gauss-Bonnet  theory. We have seen that up to the first order in derivative expansion, Gauss-Bonnet terms do not add any extra corrections to fluid entropy as expected. However, at the second order in derivative expansion, the boundary current will non-trivially depend on how we choose our horizon to boundary map, which need not be expressible entirely in terms of fluid variables. So generically, the boundary entropy current generated by dualizing the horizon current will not admit a fluid dynamical description.}

\begin{document} 
\maketitle
\flushbottom

\section{Introduction}
Black holes are interesting classical solutions in Einstein gravity which continue to exist (at least perturbatively) even if we add higher derivative corrections to the gravity action. One of the key features of black holes that makes them particularly important is that it is possible to associate an entropy with each of these solutions.
For two derivative theories of gravity, the entropy of a black hole solution could be geometrically identified with the area of the horizon - the distinct null hypersurface that shields the black hole singularity \cite{PhysRevLett.26.1344,PhysRevD.7.2333,waldbook,Hawking:1973uf,Bardeen:1973gs}. It satisfies both the first and the second law of thermodynamics. In higher derivative theories of gravity, though we know how to extend this concept of entropy \cite{WaldEnt,Iyer:1994ys} so that the first law is satisfied, we still do not know its extension to dynamical black holes so that the second law is also satisfied.
However, for  small amplitude dynamics, the entropy density could be coupled with a spatial entropy current so that together they satisfy the second law at least within this approximation \cite{Wall,our1912,our2105}. Recently this construction has been generalized to second order in amplitude expansion \cite{harveyrecent} and also in the presence of scalar fields \cite{harveyrecent,nilaymass} and gauge fields \cite{nilaymass}.

In general, it is challenging to find  dynamical black hole solutions even in Einstein gravity. One either has to use some perturbation or numerics. The perturbation in terms of the amplitude of the dynamics around a stationary solution is one such analytic technique to generate dynamical black hole solutions and as mentioned above, this is the one that has been used for the construction of the entropy density and the current on the horizon.
In this note, we would like to extend this construction of horizon entropy current to another class of dynamical black hole/brane solutions generated using derivative expansion \cite{nonlinfluid,arbitdim,Hubeny:2011hd,Rangamani:2009xk}. 

Derivative expansion is a technique that could be applied to slowly varying dynamics (not necessarily of small amplitude). In \cite{nonlinfluid}, this technique has been used to generate solutions to Einstein equations in the presence of a negative cosmological constant and in \cite{Dutta_2008}, it has been further extended to Einstein-Gauss-Bonnet theory. These solutions are asymptotically AdS and are dual to  conformal hydrodynamics with a very specific value of shear viscosity that gets corrected once the Gauss-Bonnet terms are added to the gravity action. The dual theory of hydrodynamics lives on the boundary of the AdS space, a co-dimension one hypersurface with flat metric. Such a theory of hydrodynamics always admits an entropy current - a covariant vector under the boundary Lorentz transformation, which has non-negative divergence on every solution of the fluid equations. It is natural to expect that the entropy along the dynamical horizon could be recast into one candidate for the boundary entropy current in any higher derivative theory of gravity as long as the black hole solution admits a fluid dual  (see \cite{Oz1,Oz2} for such constructions).

In the case of Einstein gravity, where the horizon area plays the role of entropy density in the black hole, one could lift the horizon entropy to the boundary by using some (non-unique) horizon-boundary map. This map finally results in an entropy current in the fluid theory, expressed entirely in terms of fluid variables and with non-negative divergence, guaranteed by the `horizon area increase' theorem \cite{entropycurrent_fluid}. In other words,  in two derivative theories of gravity with negative cosmological constant, the entropy production at every point on the dynamical horizon (with a degenerate metric) could be neatly mapped to the similar  ultra-local (point by point) entropy production in the dual fluid dynamics, living on the boundary (with simple flat metric).

Clearly, this whole algorithm of lifting the horizon entropy density to the fluid entropy current crucially depends on how we map the points on the horizon to the points on the boundary.  From the perspective of the boundary fluid, the mapping functions, which relate every point on the null horizon to a point on the time-like boundary, are some external variables. One of the key outcomes of the analysis in \cite{entropycurrent_fluid} is that for dynamical black holes/branes in Einstein gravity, it is possible to choose these mapping functions in a way so that the local entropy density on the horizon is a local function of the fluid variables only.

The reason that allows one to make such a choice is as follows.\\
For black holes in two derivative theories, the second law of thermodynamics is a consequence of the `horizon area increase' theorem. The proof of this theorem does not need any form of perturbation or approximation on the horizon dynamics \cite{PhysRevLett.26.1344,PhysRevD.7.2333,waldbook}. Also, the candidate for the entropy density -  the area of the spatial sections of the horizon, is entirely independent of  how we choose to parametrize the null generators of the horizon. This is why in two derivative theories, one is free to choose the mapping functions that are compatible with the description of the boundary fluid. 

In fact, the choice of mapping used in \cite{entropycurrent_fluid} explicitly breaks the Lorentz covariance of the boundary coordinates, and the applicability of derivative expansion is implicitly assumed at all intermediate steps. It was  the final answer for entropy current that was independently checked for Lorentz covariance and then  covariantized  entirely in terms of fluid variables and their boundary derivatives. 

Now, the construction of entropy current in \cite{our1912,our2105} in higher derivative theories depends very much on how we choose the spatial sections of the dynamical horizon. So, a priori it is not clear whether in such higher derivative theories also 
\begin{enumerate}
\item we could lift the horizon entropy current to the boundary and rewrite the entropy production as a divergence of a current covariant with respect to the boundary metric;
\item the covariant boundary entropy current, thus constructed, is a legitimate entropy current in the dual theory of hydrodynamics, expressible entirely in terms of fluid variables.
\end{enumerate}
 
In this note, we shall see that the answer to the first question is positive. We have been able to construct a manifestly covariant formula for boundary entropy current by rearranging the expressions for the entropy current and entropy density on the horizon with the mapping functions. These mapping functions are left arbitrary in our construction. They appear in the final formula of the boundary entropy current as new variables, much like the fluid variables. However, these new variables need not admit any derivative expansion.

In the case of two derivative theories of gravity, dependence on these mapping functions cancel out in the final formula as a consequence of the `reparametrization invariance' of horizon area. This provides another justification of why the procedure used in \cite{entropycurrent_fluid}, despite explicitly breaking the Lorentz invariance and translation invariance at every intermediate step,  has worked so beautifully.

But in higher derivative theories, the construction of the entropy density and the entropy current need a very specific choice of coordinates on the horizon, where the null generators are affinely parametrized.
Therefore,  unlike the two derivative theories, the mapping functions here are not completely free; they have to be compatible with the horizon adapted coordinates used in \cite{our1912,our2105} to parametrize the rate of entropy production along the null generator. Further, to generate a legitimate fluid entropy current on the boundary fluid, the mapping functions should not violate the applicability of derivative expansion in terms of the boundary coordinates. It turns out that  these two conditions are not easy to satisfy simultaneously. We applied our construction to the horizon entropy current in  Einstein-Gauss-Bonnet theory, whose fluid dual has already been constructed in \cite{Dutta_2008}. But even before using the details of the of the bulk metric here, we could see that the covariant entropy current in the boundary theory, constructed by dualizing the horizon entropy, will have non-trivial dependence on the mapping functions, which do not get cancelled and also most likely will not admit any derivative expansion.

To summarise, the answer to the second question posed above is generically negative.

However, this is probably not a complete `no go' theorem about the possibility of dualizing the `horizon entropy current' to a legitimate fluid entropy current. 
It is still possible that for some special higher derivative theory, these dependencies on the mapping functions do cancel among themselves. Also, we have one construction of the boundary entropy current, but we do not have any proof that this is a unique construction. For example, any expression of the current could be modified by adding terms that are identically conserved without affecting its divergence. Similarly, the entropy current and entropy density on the horizon also have a number of ambiguities \cite{Jacobson:1993xs,Jacobson:1993vj,Jacobson:1995uq,Reparametrization_Symmetry}. It is worth exploring whether all the terms that are not compatible with derivative expansion or fluid dynamics could be removed by fixing these ambiguities in a certain way. We leave these for future work.\\

This note is organized as follows. In the next subsection, we give a summary of the main results.
Then in section \ref{sec:coordinates} we have described how we could construct the horizon to boundary map. Next, in section \ref{sec:translation}, we have used this map to translate the horizon current to a covariant boundary current. In section \ref{sec:entcurHOR}, we have applied this construction to the dynamical black holes of Einstein-Gauss-Bonnet theory in the presence of a negative cosmological constant. In section \ref{sec:future}, we explore some future directions.
 Finally, in section \ref{sec:conclude}, we conclude.


\subsection{The Result}\label{subsec:results}

As mentioned before, the main result in this note is a formula for the boundary entropy current whose divergence is equal to the rate of local entropy production on the dynamical horizon. \\\

In \cite{our1912,our2105}, it has been shown that in higher derivative theories of gravity, one could always construct an entropy density (denoted as $j^v$) and a spatial entropy current (denoted as $j^i$) on every black hole solution with a dynamical horizon such that
\begin{equation}\label{def_j}
\left[\frac{1}{\sqrt{h}}\partial_v\left(\sqrt{h}j^v\right)+\nabla_ij^i\right]\geq0
\end{equation}

provided the amplitude of the dynamics remains small throughout the evolution of the black hole till it settles to equilibrium.\\
Here $v$ is the affine parameter along the null generators of the horizon; the sub/superscript `$i$' denotes the spatial coordinates along the constant $v$ slices of the horizon and $\nabla_i$ is the covariant derivative with respect to the induced metric along the constant $v$ slices.

In this note, using a set of mapping functions from the horizon to the boundary (a  map between the horizon coordinates $\{v,\alpha^i\}$ and boundary coordinates $\{x^\mu\}$) we have constructed an expression for entropy current $J^\mu$ on the boundary such that
\begin{equation}
D_\mu J^\mu = \frac{1}{\sqrt{h}}\partial_v (\sqrt{h} ~j^v )+ \nabla_i j^i
\end{equation}
where $D_\mu$ denotes the covariant derivative with respect to the boundary metric.\\\\
The expression for $J^\mu$ turns out to be
\begin{equation}\label{eq:jmu22}
\begin{split}
&J^\mu=\frac{1}{\sqrt{g^{(b)}}} \frac{\sqrt{H}}{\sqrt{t^\alpha t^\beta g^{(b)}_{\alpha\beta}}} \left(j^v  t^\mu + j^i l^\mu_i\right)\\
& H \equiv \frac{\hat n_\mu\hat n_\nu \epsilon^{\mu\mu_1\mu_2\cdots \mu_n}\epsilon^{\nu\nu_1\nu_2\cdots \nu_n}\chi_{\mu_1\nu_1}\cdots \chi_{\mu_n\nu_n}}{n!}\\
& \hat n^\mu = {t^\mu\over \sqrt{t^\alpha t^\beta g^{(b)}_{\alpha\beta}}}\\
\end{split}
\end{equation}
where 
$t^\mu$ and $l^\mu_i$ are vectors related to the map of $\{v,\alpha^i\}$ coordinates on the horizon to $\{x^\mu\}$ coordinates on the boundary and are defined as follows
$$t^\mu \equiv {\partial x^\mu\over \partial v},~~~l^\mu_i \equiv {\partial x^\mu\over \partial\alpha^i}$$
And $\chi_{\mu\nu}$ is the degenerate induced metric on the horizon expressed in terms of the boundary coordinates or, more precisely if the bulk metric dual to the boundary fluid is denoted as $G_{AB}(r,x^\mu)$ with $r=0$ being the horizon, then 
$$\chi_{\mu\nu}= G_{\mu\nu}\vert_{r\rightarrow 0}$$
The symbol $\epsilon^{\mu\mu_1\cdots\mu_n}$ denotes the completely antisymmetric  $(n+1)$ indexed tensor with each component equal to either $0$ or $\pm 1$. Note that in our convention, this epsilon tensor does not have any factor like the determinant of the metric.

We have explicitly constructed the boundary entropy current for the case of Einstein-Gauss-Bonnet theory, for which the horizon current is already determined in \cite{our1912}.
\begin{equation}\label{eq:curgauss}
\begin{split}
&J^\mu=\frac{1}{\sqrt{g^{(b)}}} {\sqrt{H}\over \sqrt{t^\alpha t^\beta g^{(b)}_{\alpha\beta}}}\bigg[ \left(1+\alpha^2 {\cal R}\right)  t^\mu -4\alpha^2\left(\bar\chi^{\gamma\alpha}\bar\chi^{\mu\beta} - \bar\chi^{\gamma\mu}\bar\chi^{\alpha\beta}\right)\left({\cal D}_\gamma{\cal K}_{\alpha\beta}\right)\bigg]\\
&\text{with the following notation}\\
&{\cal R} \equiv \left(\bar\chi^{\mu_1\nu_1}\bar\chi^{\mu_2\nu_2}-\bar\chi^{\mu_1\nu_2}\bar\chi^{\mu_2\nu_1}\right)\bigg[\partial_{\mu_1}\Gamma_{\nu_1,\mu_2\nu_2} - \bar\chi^{\alpha_1\alpha_2}\Gamma_{\alpha_1,\mu_1\nu_1}\Gamma_{\alpha_2,\mu_2\nu_2}-2 t^\alpha\Gamma_{\alpha,\mu_1\nu_1}\left(\partial_{\mu_2}\tilde t_{\nu_2}\right)\bigg]_{r=0}\\
&{\cal K}_{\alpha\beta} \equiv -t^\mu\Gamma_{\mu,\alpha\beta},~~~\bar\chi^{\mu\nu} \equiv \left(\delta_\alpha^\mu - t^\mu\tilde t_\alpha\right)\left(\delta_\beta^\nu - t^\nu\tilde t_\beta\right)G^{\alpha\beta}(r=0)\\
&\text{and}\\
&{\cal D}_\alpha{\cal K}_{\mu\nu} \equiv \partial_\alpha{\cal K}_{\mu\nu} - \tilde\Gamma^\beta_{~\alpha\mu} {\cal K}_{\beta\nu} - \tilde\Gamma^\beta_{~\alpha\nu} {\cal K}_{\mu\beta}
\end{split}
\end{equation}
where
\begin{equation}\label{eq:nota}
\begin{split}
&\tilde t_\mu \equiv {\partial v\over\partial x^\mu} ~~\text{such that}~~t^\mu\tilde t_\mu =1,~~l^\mu_i \tilde t_\mu =0\\
&\Gamma_{\alpha,\mu\nu} ={1\over 2}\left(\partial_\mu \chi_{\nu\alpha}+\partial_\nu \chi_{\mu\alpha}- \partial_\alpha \chi_{\mu\nu}\right),~~~\tilde\Gamma^\alpha_{\mu\nu}\equiv\bar\chi^{\alpha\beta}\Gamma_{\beta,\mu\nu}+ t^\alpha\partial_\nu\tilde t_\mu \\
\end{split}
\end{equation}
Note that for a generic case, these mapping functions will enter  the expression of the boundary entropy current through the two vectors $t^\mu$ and $\tilde t_\mu$. And as we have mentioned before, these two vectors need not admit a derivative expansion.
The reason is as follows.\\
$t^\mu$, being  the tangent vector to the affinely parametrized null generators  of the horizon (located at $r=0$), must be proportional to the normal  of the $r=0$ hypersurface. This normal  is given by $n^\mu\equiv G^{\mu  r}(r=0)$, which according to fluid-gravity correspondence, must admit a derivative expansion in terms of fluid variables. Let us denote the proportionality factor as $e^{\phi(x)}$.
$$t^\mu = e^{\phi(x)} n^\mu =\ e^{\phi(x)}~ G^{\mu r}\vert_{r=0}$$
The affine parameter $v$ could be related to $\phi(x)$ as (see section \ref{subsec:enthigh})
$$v \equiv e^{-\phi} L = e^{-\phi}\sum_{k=0}^\infty L_{(k)},~~\text{where}~~{L_{(k)}\over L_{(0)}}= -\left[(n\cdot\partial)L_{(k-1)}\right],~~~L_{(0)} = -\left[ \left(n^\mu n^\nu\over 2\right)\left[ \partial_r \chi_{\mu\nu}\right]_{r=0}\right]^{-1}$$
Therefore $\tilde t_\mu = \left(\partial v\over \partial x^\mu\right)$ must have a term proportional to $\partial_\mu\phi$.
Now $\partial_\mu\phi$ must be a zeroth order vector since its component along the direction of $n^\mu$ is of zeroth order in derivative expansion. It satisfies the equation (follows from the fact that $t^\mu$ is an affinely parametrized geodesic, see section \ref{subsec:CoordBoundary})
$$(n\cdot\partial)\phi= \left(n^\mu n^\nu\over 2\right)\left[ \partial_r \chi_{\mu\nu}\right]_{r=0}$$

However at zeroth order in derivative expansion, only vector that could be expressed entirely in terms of fluid variables is the fluid velocity $u^\mu$ itself.
So $\partial_\mu\phi$ has to be proportional to $u_\mu$ with proportionality factor being some function of temperature. But any gradient vector field like $\partial_\mu\phi$ or $\partial_\mu v$ could not be proportional to fluid velocity whenever the velocity has nonzero vorticity. This shows that any generic situation $\partial_\mu\phi$ are the `non-fluid' terms, that will remain there in the boundary entropy current constructed dualizing the horizon current.\\

Finally we have evaluated the boundary current \eqref{eq:jmu2} on slowly varying black holes in Einstein-Gauss-Bonnet theory up to correction of order ${\cal O}(\partial^2)$. Up to this order in derivative expansion, the `non-fluid' mapping functions (functions that do not admit a derivative expansion in terms of fluid variables) do not contribute. In fact just like the fluid dual to Einstein gravity, the ${\cal O}(\partial)$ contribution to the entropy current vanishes which is also what is expected for an uncharged fluid. The expression of $J^\mu$ turns out to be the following
\begin{equation}\label{eq:Jmuexp}
\begin{split}
{\color{black}J^\mu=r_H^3~\hat{n}^\mu+\mathcal{O}\left(\alpha^4,\partial^2\right)}\\
\end{split}
\end{equation}
where $r_H$ is the length scale associated with the temperature of the Black hole or the dual fluid as  defined in \eqref{fluidmetric}.


\section{The map between horizon and boundary}\label{sec:coordinates}
As described before, the dynamical black brane solution that we are considering here, is always perturbative. Two different types of perturbations are used to describe the solution. For the entropy density and the current constructed on the horizon as in \cite{Wall,our1912,our2105,harveyrecent}, the perturbation parameter is the amplitude of the dynamics whereas in \cite{nonlinfluid} it is the derivatives of the boundary fluid data (veocity and temperature) that play the role of the small parameter. In both the cases the starting point is a stationary  black hole/brane metric. In both cases we could choose a gauge where the horizon is at the origin of the radial coordinate (the coordinate that measures the distance away from the horizon). In amplitude expansion the black hole metric is parametrized by its components evaluated at the horizon whereas in case of derivative expansion it is parametrized by the metric components evaluated at the AdS boundary expressed in terms of the variables of the dual fluid description.

The key part of this note is about a map between the points on the horizon and the points on the boundary.
To define any such map we first need to set up coordinate systems on both horizon and the boundary. In this section first we shall briefly describe the two coordinate systems that are used to describe the entropy current on the horizon \cite{our1912,our2105} and the fluid dynamics living on the boundary \cite{nonlinfluid,arbitdim}. We shall refer to them as `horizon adapted coordinates' and `boundary coordinates' respectively.\\
Then in the final subsection we shall relate this two coordinates to get a point by point map from the horizon to the boundary.

\subsection{Horizon adapted coordinate system}\label{subsec:CoordHorizon}
 The entropy density and current, defined on the horizon are expressed in a very special choice of coordinates, tuned to the structure of null hypersurface. We shall denote this coordinate system as `horizon adapted coordinate system'. In these coordinates the metric takes the following form
\begin{equation}\label{eqLhorizonmet}
\begin{split}
ds^2 = 2 d\rho~dv - \rho^2X(\rho,v,\vec \alpha)~dv^2 +2\rho~\omega_i(\rho,v,\vec \alpha) ~dv~d\alpha^i + h_{ij}(\rho,v,\vec \alpha) ~d\alpha^id\alpha^j
\end{split}
\end{equation}
where $X$, $\omega_i$ and $h_{ij}$ are arbitrary nonzero functions of $\rho,v$ and $\vec \alpha = \{\alpha^i\}$. In this metric, the horizon is located at the $\rho=0$ hypersurface.
At $\rho=0$, the vector $\partial_v $ is affinely patrametrized null generator of the horizon, with $v$ being the affine parameter. $\partial_i$ s are the spatial vectors on the constant $v$ slices of the horizon. The entropy current is defined on the horizon and therefore could depend only on the metric functions $X$, $\omega_i$ and $h_{ij}$ and their $\partial_i$ and $\partial_v$ derivatives. \\

In a stationary solution, the $\rho$ and $v$ dependence of the metric would be constrained. The functions $X$, $\omega_i$ and $h_{ij}$ will only depend on the product of $\rho$ and $v$. The stationary metric will be completely invariant under the transformation
$$v\rightarrow \lambda v,~~~~\rho\rightarrow{\rho\over \lambda}$$
While constructing the horizon entropy current, a departure from this invariance has been treated as the small parameter, characterizing the amplitude of the dynamics.


\subsection{Boundary coordinates}\label{subsec:CoordBoundary}

In hydrodynamics, the local velocity of the field  denoted as $u^\mu(x)$ is a special vector. While writing the dual metric, the most convenient choice of gauge turns out to be related to this velocity field. In this choice of gauge (with coordinates denoted as $\{{ r}, y^\mu\}$), the metric takes the following general structure
\begin{equation}\label{fluidmetric}
\begin{split}
&ds^2 = -2 u_\mu ~dy^\mu~d r +{\chi}_{\mu\nu} ~dy^\mu dy^\nu\\
&\text{$\chi_{\mu\nu}$ could be further decomposed as }\\
&\chi_{\mu\nu} \equiv  S_1 u_\mu u_\nu + S_2 P_{\mu\nu} + ( V_\mu u_\nu + V_\nu u_\mu)+ {\cal T}_{\mu\nu}\\
\text{such that} ~~&u^\mu V_\mu = u^\mu{\cal T}_{\mu\nu}=0,~~~P_{\mu\nu} \equiv \eta_{\mu\nu} + u_\mu u_\nu
\end{split}
\end{equation}
Here $ r\rightarrow\infty$ is the boundary, and the metric takes the form of Poincare patch AdS as we approach the boundary. Here also, we shall choose the origin of the $r$ coordinate at the horizon. Therefore $r=0$ is a null hypersurface by construction, which further implies
$$G^{rr} (r=0) =0 ~~\text{and}~~n^\mu\partial_\mu = G^{r\mu}(r=0)\partial_\mu~~\text{is a null vector at the horizon}$$
The vector 
$n^\mu= G^{r\mu}\vert_{r=0}$ must be identified with the null generator of the horizon (though not affinely parametrized).\\
Using the fact that the null generator of the horizon is just the dual vector of the one form $dr$ or, in other words, $n^A G_{AB} =\delta^r_B$, we get the following identities for the $n^\mu$ vector, which would turn out to be useful at a later point.
\begin{equation}\label{id0}
\begin{split}
&\delta^r_B= n^A G_{AB}\vert_{r=0}
=n^\mu G_{\mu B}\\
\Rightarrow~& n^\mu G_{\mu r} = - u_\mu n^\mu = 1\\
\Rightarrow~&n^\mu G_{\mu\nu}\vert_{r=0} = n^\mu\chi_{\mu\nu}\vert_{r=0} =0
\end{split}
\end{equation}
 \\
$ S_1$, $S_2$, $ V_\mu$ and ${\cal T}_{\mu\nu}$ all are functions of $r$ and $y^\mu$ , but the $y^\mu$ dependence is known only perturbatively where the perturbation parameters are the derivatives of the fluid variables. In fact the derivative expansion would be valid only when the fluid variables are slowly varying with respect to some scale, in this case, the temperature of the fluid. The more the number of derivatives, the more suppressed the terms are. \footnote{Note in \cite{nonlinfluid,Dutta_2008} the choice of gauge was quite different from the one we are using here. In case of fluid gravity correspondence, it makes sense to parametrize the metric in terms of fluid variables defined with respect to the boundary stress tensor. The horizon in the initial papers of fluid-gravity correspondence is not located at $ r=0$ but given by $ r= r_H(y^\mu)$ whose value is related to the local temperature of the dynamical black brane being considered. We  can translate between these two gauges by a simple shift of $r$ coordinate
$$r\rightarrow  r +r_H(y^\mu)$$
 This step adds a little modification to the fluid metric without affecting its general structure. The net result of this shift is just a shift in $\chi_{\mu\nu}$ as follows
\begin{equation}\label{shift}
\begin{split}
&\chi_{\mu\nu}\rightarrow  \chi_{\mu\nu} - (u_\mu\partial_\nu+u_\nu\partial_\mu) r_H\\
\end{split}
\end{equation}
In our solution $r_H$ will simply be length scale, with respect to which the slow variation or the derivative expansion is defined.}


\subsection{The horizon to boundary map}\label{sec:MapHorBoundary}
The metric described in section \ref{subsec:CoordHorizon} is in a completely different gauge than that of hydrodynamic metric in section \ref{subsec:CoordBoundary}. However, the construction of the horizon entropy current is very much tied to the choice of coordinates as given in \ref{subsec:CoordHorizon}. It is obvious that, to translate the horizon entropy current in terms of the fluid variables the first step would be to establish a dictionary between these two coordinate systems. \\
We shall transform the fluid metric (as given in eq:\eqref{fluidmetric}) to the gauge described in section (\ref{subsec:CoordHorizon}). This will allow us to describe metric functions ($X$, $\omega_i$ and $h_{ij}$) as they appeared in equation \eqref{eqLhorizonmet} in terms of the fluid variables (velocity and temperatures) and their appropriate derivatives. 

In other words, we shall express $r$ and $x^\mu$ as functions of $\{\rho,v,\vec\alpha\}$ such that the following gauge conditions are satisfied.
\begin{equation}\label{gauge-conditions}
\begin{split}
G_{\rho\rho}=0\Rightarrow~&-2 u_\mu \left(\partial x^\mu\over \partial\rho\right) \left(\partial r\over \partial\rho\right) + \chi_{\mu\nu}\left(\partial x^\mu\over \partial\rho\right) \left(\partial x^\nu\over \partial\rho\right) =0\\
G_{\rho v}=1\Rightarrow~&-u_\mu \left[\left(\partial r\over \partial \rho\right)\left(\partial x^\mu\over \partial v\right) + \left(\partial r\over \partial v\right)\left(\partial x^\mu\over \partial\rho\right)\right] +\chi_{\mu\nu} \left[\left(\partial x^\nu\over \partial \rho\right)\left(\partial x^\mu\over \partial v\right)\right]=1\\
G_{\rho\alpha_i} =0\Rightarrow~&-u_\mu \left[\left(\partial r\over \partial \rho\right)\left(\partial x^\mu\over \partial\alpha_i\right) + \left(\partial r\over \partial \alpha_i\right)\left(\partial x^\mu\over \partial\rho\right)\right] +\chi_{\mu\nu} \left[\left(\partial x^\nu\over \partial \rho\right)\left(\partial x^\mu\over \partial\alpha_i\right)\right]=0
\end{split}
\end{equation}

Now it is difficult to solve these equations exactly, even in just the radial coordinate. However, for our entropy current, it is enough to have the near horizon structure of the metric. So we shall be solving the gauge conditions \eqref{gauge-conditions} in an expansion in $\rho$.

We shall take the following ansatz for the  coordinate transformations:
\begin{equation}\label{coordi-1}
\begin{split}
r &= \rho ~r_{(1)}(v,\alpha_i) + \rho^2 ~r_{(2)}(v,\alpha_i) + \cdots\\
x^\mu &= x^\mu_{(0)}(v,\alpha_i)+\rho ~x^\mu_{(1)}(v,\alpha_i) + \rho^2 ~x^\mu_{(2)}(v,\alpha_i) + \cdots\\
\end{split}
\end{equation}
In the above coordinate transformation the functions $x^\mu_{(0)}(v,\alpha^i) $ will be effectively taken as input functions. All the rest, namely $\{x^\mu_{(n)}(v,\alpha^i)\}$ and $r_{(n)}(v,\alpha^i)$ will be determined in terms of the functions $x^\mu_{(0)}(v,\alpha_i)$. In Appendix \ref{transformation_functions} we have determined the first few coefficients of the above transformation equations (equation \eqref{coordi-1}).\\
Note that the input functions $x^\mu_{(0)}(v,\alpha_i)$ are not entirely free. The vector 
$t^\mu \equiv \left(\partial x^\mu_{(0)}\over \partial v\right)$ must be an affinely parametrized null geodesic with respect to the full metric.\\
Let us define the following set of vectors that are tangent to the horizon
$$t^\mu \equiv \left(\partial x^\mu_{(0)}\over \partial v\right),~~l^\mu_i \equiv\left(\partial x^\mu_{(0)}\over \partial\alpha_i\right)$$
$t^\mu$, being  the null generator of the horizon, is also a normal to the horizon. \\
Hence it follows that $t^\mu$ must be proportional to $n^\mu$ of the fluid metric we defined in the previous step. In other words
$$t^\mu = e^\phi n^\mu = e^\phi G^{r\mu}(r=0)$$
where $\phi$ is a scalar function of $\{x^\mu\}$ so that $t^\mu$  becomes a affinely parametrized null geodesic. Processing this condition we get the following equation for the field $\phi(x)$
\begin{equation}\label{phieqn}
\begin{split}
(n\cdot\partial)\phi= \left(n^\mu n^\nu\over 2\right)\left[ \partial_r \chi_{\mu\nu}\right]_{r=0}
\end{split}
\end{equation}
Note that the RHS of equation \eqref{phieqn} is nonzero even at zeroth order in derivative. Therefore, it is not $\phi$ but its derivative along the direction of $n^\mu$ that satisfies the derivative expansion. At this stage we are free to choose the dependence of $\phi$ along the directions perpendicular to $n^\mu$.

Now $\phi$ is an external scalar field from the perspective of boundary fluid dynamics and generically the fluid entropy current would depend on the choice of $\phi$. We should be able to choose $\phi$ in a way so that the final fluid entropy current is entirely expressible in terms of the fluid variables like velocity and temperature only.

\section{Translating the horizon current to the boundary current}\label{sec:translation}
In this section, we shall  find out an abstract expression for entropy current $J^\mu$ in the boundary such that 
\begin{equation}\label{eq:goalhere}
D_\mu J^\mu (j^v, j^i) = {1\over\sqrt{h}}\partial_v (\sqrt{h} ~j^v )+ \nabla_i j^i
\end{equation}
where $j^v$ and $j^i$ are defined in equation \eqref{def_j}. Here the RHS of the above equation is written in the horizon adapted coordinates whereas the LHS is in terms of the boundary coordinates. $D_\mu$ denotes the covariant derivative with respect to the boundary metric.  In the first subsection, we shall describe how to determine $J^\mu$, given $j^v$ and $j^i$. The final expression for $J^\mu$ turns out to be the following
\begin{equation}\label{eq:jmu}
J^\mu= \frac{1}{\sqrt{g^{(b)}}}\sqrt{H} \left(j^v t^\mu +j^i l^\mu_i\right)~~\text{where}~~ H \equiv \frac{\tilde t_\mu\tilde t_\nu \epsilon^{\mu\mu_1\mu_2\cdots \mu_n}\epsilon^{\nu\nu_1\nu_2\cdots \nu_n}\chi_{\mu_1\nu_1}\cdots \chi_{\mu_n\nu_n}}{n!}
\end{equation}
Here $j^v$ and $j^i$ have to be read off from the expression of the horizon current.
$t^\mu$, $\tilde t_\mu$ and $l^\mu_i$ are vectors related to the map.
$$t^\mu \equiv {\partial x^\mu\over \partial v},~~\tilde t_\mu \equiv {\partial v\over \partial x^\mu},~~~l^\mu_i \equiv {\partial x^\mu\over \partial\alpha^i},~~~l^i_\mu \equiv {\partial \alpha^i\over \partial x^\mu}$$
Using the fact that $t^\mu = e^\phi n^\mu $ and $t^\mu \tilde t_\mu =1$ the expression for $H$ and current could be simplified further
\begin{equation}\label{eq:jmu2}
\begin{split}
&J^\mu=\frac{1}{\sqrt{g^{(b)}}} \frac{\sqrt{H}}{\sqrt{t^\alpha t^\beta g^{(b)}_{\alpha\beta}}} \left(j^v  t^\mu + j^i l^\mu_i\right)\\
& H \equiv \frac{\hat n_\mu\hat n_\nu \epsilon^{\mu\mu_1\mu_2\cdots \mu_n}\epsilon^{\nu\nu_1\nu_2\cdots \nu_n}\chi_{\mu_1\nu_1}\cdots \chi_{\mu_n\nu_n}}{n!}\\
& \hat n^\mu = {n^\mu\over \sqrt{n^\alpha n^\beta g^{(b)}_{\alpha\beta}}}= {t^\mu\over \sqrt{t^\alpha t^\beta g^{(b)}_{\alpha\beta}}}\\
\end{split}
\end{equation}

\subsection{Constructing $J^\mu$}
In this  subsection, we shall determine an algorithm to determine $J^\mu$ out of $j^v$ and $j^i$. The  key issue here is to re-express the entropy production formula on the horizon (i.e., the expression in the RHS of equation \eqref{eq:goalhere}) as a divergence of a current covariant with respect to the boundary coordinates. It turns out if we could rewrite the equation \eqref{eq:goalhere} in  a `metric independent' language using $n$ and $(n-1)$ forms, it helps to identify the $J^\mu$. \\
Let us first define the following two $n$-forms.
\begin{equation}\label{eq:form1}
\begin{split}
&J_{temp}\equiv \sqrt{h}~j^v~ {\epsilon_{i_1i_2\cdots i_n}d\alpha^{i_1}\wedge d\alpha^{i_2}\cdots\wedge d\alpha^{i_n}\over n!}\\
&J_{space}\equiv -\sqrt{h}~j^k~ {\epsilon_{ki_2i_3\cdots i_n} dv\wedge d\alpha^{i_2}\wedge d\alpha^{i_3}\cdots\wedge d\alpha^{i_n}\over (n-1)!}\\
\end{split}
\end{equation}
Here, $\epsilon_{i_1i_2\cdots i_n}$ is the completely antisymmetric $n$ indexed tensor with each component equal to $0$ or $\pm 1$. 

One could show that the  exterior derivative of $(J_{temp} + J_{space})$  is proportional to the top form on the horizon,  where the proportionality constant is the RHS of equation \eqref{eq:goalhere}.
\begin{equation}\label{eq:form2}
d(J_{temp} + J_{space})=\left[\partial_v\left(\sqrt h~j^v\right) + \sqrt{h}~\nabla_i j^i \right] \left[{\epsilon_{i_1i_2\cdots i_n}dv\wedge d\alpha^{i_1}\wedge d\alpha^{i_2}\cdots\wedge d\alpha^{i_n}\over n!}\right]
\end{equation}
Here $d$ denotes the exterior derivative.\\
Now we shall rewrite  $J_{temp}$ and $J_{space}$ in terms of the boundary coordinates using the fact that 
$$dv = \tilde t _\mu dx^\mu,~~~d\alpha^i = l^i_\mu dx^\mu$$
We need to use the following identities.
\begin{equation}\label{eq:identities}
\begin{split}
&l^{i_1}_{\mu_1}\cdots l^{i_n}_{\mu_n} \epsilon_{i_1i_2\cdots i_n} = \Delta~ t^\mu~\epsilon_{\mu\mu_1\cdots\mu_n} \\
&l_{i_1}^{\mu_1}\cdots l_{i_n}^{\mu_n} \epsilon^{i_1i_2\cdots i_n} = \left(1\over\Delta\right)~ \tilde t_\mu~\epsilon^{\mu\mu_1\cdots\mu_n} \\
&\epsilon_{\mu\mu_1\mu_2\cdots\mu_n}\left( dx^\nu\wedge dx^{\mu_1}\cdots\wedge dx^{\mu_n}\over n!\right)=\delta^\nu_\mu \epsilon_{\mu_1\cdots\mu_{n+1}} \left( dx^{\mu_1}\wedge \cdots\wedge dx^{\mu_{n+1}}\over (n+1)!\right)\\
&\epsilon_{\mu\alpha\mu_2\cdots\mu_n}\left( dx^\nu\wedge dx^{\mu_2}\cdots\wedge dx^{\mu_n}\over (n-1)!\right)=\left(\delta^\nu_\alpha \epsilon_{\mu\mu_1\cdots\mu_n} -  \delta^\nu_\mu \epsilon_{\alpha\mu_1\cdots\mu_n}  \right)\left( dx^{\mu_1}\wedge \cdots\wedge dx^{\mu_n}\over n!\right)\\
\end{split}
\end{equation}
Here $\Delta$ is the Jacobian of the coordinate transformation
$$\Delta = det\left[\partial\{v,\alpha^i\}\over \partial\{ x^\mu\}\right],~~~{1\over \Delta} = det\left[\partial\{ x^\mu\}\over \partial\{v,\alpha^i\}\right],$$
First, we shall write an expression for $\sqrt h$ in terms of the boundary coordinates. 
\begin{equation}\label{eq:deth}
\begin{split}
h = det[h_{ij}] &={ \epsilon^{i_1\cdots i_n}\epsilon^{j_1\cdots j_n}h_{i_1j_1}\cdots h_{i_nj_n}\over n!}\\
& = \left(1\over n!\right)\epsilon^{i_1\cdots i_n}\epsilon^{j_1\cdots j_n}\left[l^\mu_{i_1}\cdots l^{\mu_n}_{i_n}\right]\left[l^{\nu_1}_{j_1}\cdots l^{\nu_n}_{j_n}\right]\chi_{\mu_1\nu_1}\cdots \chi_{\mu_n\nu_n}\\
&=\left(1\over n!\right)\left(1\over \Delta\right)^2\tilde t_\mu\tilde t_\nu~\epsilon^{\mu\mu_1\cdots \mu_n}\epsilon^{\nu\nu_1\cdots \nu_n} ~\chi_{\mu_1\nu_1}\cdots \chi_{\mu_n\nu_n}\\
\end{split}
\end{equation}
Using the above identities we process both $J_{temp}$ and $J_{space}$ as follows.
\begin{equation}\label{eq:formtemp}
\begin{split}
J_{temp} &= \sqrt{h}~ j^v \left(\epsilon_{i_1\cdots i_n} l^{i_1}_{\mu_ 1}\cdots l^{i_n}_{\mu_n} \over n!\right)~dx^{\mu_1}\wedge\cdots \wedge dx^{\mu_n}\\
&=  \sqrt{h}\Delta~ j^v \left(t^\mu\epsilon_{\mu\mu_1\cdots \mu_n}\over n!\right)~dx^{\mu_1}\wedge\cdots \wedge dx^{\mu_n}\\
&= \left[\tilde t_\alpha\tilde t_\beta~\epsilon^{\alpha\alpha_1\cdots \alpha_n}\epsilon^{\beta\beta_1\cdots \beta_n} ~\chi_{\alpha_1\beta_1}\cdots \chi_{\alpha_n\beta_n}\over n!\right]^{1\over 2} j^v \left(t^\mu\epsilon_{\mu\mu_1\cdots \mu_n}\over n!\right)~dx^{\mu_1}\wedge\cdots \wedge dx^{\mu_n}\\
\end{split} 
\end{equation}

\begin{equation}\label{eq:formspace}
\begin{split}
J_{space} &= -\sqrt{h} ~j^k\left(\epsilon_{ki_2i_3\cdots i_n} \over (n-1)!\right)\tilde t_\mu l^{i_2}_{\mu_2}\cdots l^{i_n}_{\mu_n}~dx^\mu\wedge dx^{\mu_2}\cdots\wedge dx^{\mu_n}\\
 &= -\sqrt{h} ~\left(j^k l^{\nu}_k\right)\left(\epsilon_{i_1i_2i_3\cdots i_n} \over (n-1)!\right)\tilde t_\mu l^{i_1}_{\nu}l^{i_2}_{\mu_2}\cdots l^{i_n}_{\mu_n}~dx^\mu\wedge dx^{\mu_2}\cdots\wedge dx^{\mu_n}\\
  &= -\sqrt{h} ~\left(j^k l^{\nu}_k\right)\tilde t_\mu\left(\Delta \over (n-1)!\right)t^\alpha\epsilon_{\alpha\nu\mu_2\cdots\mu_n}~dx^\mu\wedge dx^{\mu_2}\cdots\wedge dx^{\mu_n}\\
  &= -\sqrt{h}\left(\Delta \over n!\right) ~\left(j^k l^{\nu}_k\right)\tilde t_\mu t^\alpha\left[\delta^\mu_\nu\epsilon_{\alpha\mu_1\mu_2\cdots\mu_n}-\delta^\mu_\alpha\epsilon_{\nu\mu_1\mu_2\cdots\mu_n}\right]~dx^{\mu_1}\wedge dx^{\mu_2}\cdots\wedge dx^{\mu_n}\\
  &=\left[\tilde t_\alpha\tilde t_\beta~\epsilon^{\alpha\alpha_1\cdots \alpha_n}\epsilon^{\beta\beta_1\cdots \beta_n} ~\chi_{\alpha_1\beta_1}\cdots \chi_{\alpha_n\beta_n}\over n!\right]^{1\over 2} \left(\frac{(j^k l^\mu_k)~\epsilon_{\mu\mu_1\cdots \mu_n}}{n!}\right) ~dx^{\mu_1}\wedge dx^{\mu_2}\cdots\wedge dx^{\mu_n}\\
\end{split}
\end{equation}

So finally we have

\begin{equation}\label{eq:jtimeplusspace}
\begin{split}
&J_{temp} + J_{space} = \left[\tilde t_\alpha\tilde t_\beta~\epsilon^{\alpha\alpha_1\cdots \alpha_n}\epsilon^{\beta\beta_1\cdots \beta_n} ~\chi_{\alpha_1\beta_1}\cdots \chi_{\alpha_n\beta_n}\over n!\right]^{1\over 2} \left[j^v t^\mu + (j^k l^\mu_k)\right]~\frac{\epsilon_{\mu\mu_1\cdots \mu_n}}{n!} ~dx^{\mu_1}\wedge dx^{\mu_2}\cdots\wedge dx^{\mu_n}\\
\\
\Rightarrow~&d(J_{temp} + J_{space} )\\
=~&\partial_\mu\bigg( \left[\tilde t_\alpha\tilde t_\beta~\epsilon^{\alpha\alpha_1\cdots \alpha_n}\epsilon^{\beta\beta_1\cdots \beta_n} ~\chi_{\alpha_1\beta_1}\cdots \chi_{\alpha_n\beta_n}\over n!\right]^{1\over 2} \left[j^v t^\mu + (j^k l^\mu_k)\right]\bigg)\left({\epsilon_{\nu_1\cdots\nu_{n+1}}dx^{\nu_1}\wedge\cdots \wedge dx^{\nu_{n+1}}\over (n+1)!}\right)
\end{split}
\end{equation}
Now from equation \eqref{eq:form2} we know the expression of $d(J_{temp} + J_{space})$ in terms of $\{v,\alpha^i\}$ coordinate system. If we rewrite the $(n+1)$ form that appears in equation \eqref{eq:form2} in terms of $\{x^\mu\}$ coordinates we get the following
\begin{equation}\label{eq:formrewrite}
\begin{split}
&\epsilon_{i_1\cdots i_n} \left(dv\wedge d\alpha^{i_1}\cdots d\alpha^{i_n}\over n!\right)\\
=~&\tilde t_\nu~ l^{i_1}_{\mu_1}\cdots l^{i_n}_{\mu_n}\epsilon_{i_1\cdots i_n} \left(dx^\nu\wedge dx^{\mu_1}\wedge\cdots \wedge dx^{\mu_n}\over n!\right)\\
=~&\Delta~\tilde t_\nu~t^\mu \epsilon_{\mu\mu_1\cdots\mu_n} \left(dx^\nu\wedge dx^{\mu_1}\wedge\cdots \wedge dx^{\mu_n}\over n!\right)\\
=~&\Delta~\tilde t_\nu~t^\mu \delta^\nu_\mu\epsilon_{\mu_1\mu_2\cdots\mu_{n+1}} \left(dx^{\mu_1}\wedge\cdots \wedge dx^{\mu_{n+1}}\over (n+1)!\right)\\
=~&\Delta~\epsilon_{\mu_1\mu_2\cdots\mu_{n+1}} \left(dx^{\mu_1}\wedge\cdots \wedge dx^{\mu_{n+1}}\over (n+1)!\right)
\end{split}
\end{equation}

Substituting equation \eqref{eq:formrewrite} in equation \eqref{eq:form2} and then comparing with equation \eqref{eq:jtimeplusspace} we find
\begin{equation}\label{eq:fincompare}
\begin{split}
\Delta\left[\partial_v\left(\sqrt h~j^v\right) + \sqrt{h}~\nabla_i  j^i \right]= \partial_\mu\bigg( \left[\tilde t_\alpha\tilde t_\beta~\epsilon^{\alpha\alpha_1\cdots \alpha_n}\epsilon^{\beta\beta_1\cdots \beta_n} ~\chi_{\alpha_1\beta_1}\cdots \chi_{\alpha_n\beta_n}\over n!\right]^{1\over 2} \left[j^v t^\mu + (j^k l^\mu_k)\right]\bigg)
\end{split}
\end{equation}

As we have discussed before, in dynamical black holes, the expression $\left[\partial_v\left(\sqrt h~j^v\right) + \sqrt{h}~\nabla_i  j^i \right]$ is identified with net entropy production in every infinitesimal subregion of the horizon and, up to the linear order in the amplitude of the dynamics, it must vanish (if it does not, then  the same expression at linear order, will lead to both entropy production and destruction depending on the sign of the amplitude  and thus violating the second law). Since $\Delta$,  the Jacobian of the coordinate transformation is  non-vanishing everywhere, we conclude
$$\partial_\mu\bigg( \left[\tilde t_\alpha\tilde t_\beta~\epsilon^{\alpha\alpha_1\cdots \alpha_n}\epsilon^{\beta\beta_1\cdots \beta_n} ~\chi_{\alpha_1\beta_1}\cdots \chi_{\alpha_n\beta_n}\over n!\right]^{1\over 2} \left[j^v t^\mu + (j^k l^\mu_k)\right]\bigg) = 0 ~~~~\text{(up to terms nonlinear in amplitude)}$$

Now we can turn the above expression into a divergence of current covariant (i.e., in the form of equation \eqref{eq:goalhere}) with respect to the boundary metric, if we identify the boundary entropy current as
\begin{equation}\label{eq:boundaryentcur}
\begin{split}
J^\mu = {1\over\sqrt{g^{(b)}}}\bigg( \left[\tilde t_\alpha\tilde t_\beta~\epsilon^{\alpha\alpha_1\cdots \alpha_n}\epsilon^{\beta\beta_1\cdots \beta_n} ~\chi_{\alpha_1\beta_1}\cdots \chi_{\alpha_n\beta_n}\over n!\right]^{1\over 2} \left[j^v t^\mu + (j^k l^\mu_k)\right]\bigg)
\end{split}
\end{equation}
where $\bigg[g^{(b)}_{\mu\nu}= \lim_{r\rightarrow\infty}\left(\chi_{\mu\nu}\over r^2\right)\bigg]$ is the boundary metric and $g^{(b)} = det[g^{(b)}_{\mu\nu}]$.

Equation \eqref{eq:boundaryentcur} is one of our key results. 
Now a couple of comments about this formula.
\begin{itemize}
\item $J^\mu$ is a covariant vector in the boundary spacetime with boundary metric $g^{(b)}_{\mu\nu}$, provided we treat $t^\mu$, $l^\mu_k$ and $\tilde t_\mu$ as independent upper and lower index vectors respectively.

\item  Though we have said that $t^\mu$ is the affinely parametrized null generator on the horizon expressed in terms of boundary coordinates, the analysis in this section nowhere used the affineness of the $v$ parameter. So equation \eqref{eq:boundaryentcur} is valid even when $v$ is not an affine parameter, but it has to be a parameter along the null generator\footnote{For example, in \cite{entropycurrent_fluid} the null generators are parametrized using the boundary time-like coordinate $v$. This is not an affine parametrization, but still will we could apply our formula to recover the expression of entropy current derived in \cite{entropycurrent_fluid}. We have to use the following facts. In two derivative theories $j^v = 1,~~~j^i =0$ and the choice of map in \cite{entropycurrent_fluid} is such that
$\tilde t_\mu dx^\mu = dv,~~~t^\mu = {n^\mu\over n^v}$. The boundary metric $g^{(b)}_{\mu\nu}= \eta_{\mu\nu}$. } .

\item The expressions for $j^v$ and $j^i$ depend on the details of the equation of motion in  higher derivative theory, which in turn depend on the affine parametrization of the null generators.

\item $j^v$ and $j^i$ could be determined in terms of the functions appearing in metric \eqref{eqLhorizonmet}  (i.e., $X$, $\omega_i$ and $h_{ij}$) and their appropriate derivatives. Using this horizon to boundary map, we could re-express $j^v$ and $j^i$ in terms of the fluid variables and the mapping vectors $t^\mu$ and  $l^\mu_i$.
\item From the perspective of boundary fluid, $t^\mu$, $l^\mu_k$ or $\tilde t_\mu$ are external variables. So the entropy current described in equation \eqref{eq:boundaryentcur} would be a genuine fluid entropy current provided  our mapping functions are such that the vectors  $t^\mu$, $l^\mu_k$ or $\tilde t_\mu$ are either constants or are determined entirely in terms of fluid variables.
\end{itemize}

\subsection{Entropy current in boundary fluid dual to Einstein gravity }\label{boundary_current_Einstein}
In Einstein gravity, the entropy on the horizon is simply given by the area of the spatial sections of the horizon. In our choice of horizon-adapted coordinate system, it is the square root of the determinant of $h_{ij}$. It follows
$$j^v_{(2)} =1,~~~j^i_{(2)}=0$$
where the subscript ${(2)}$ denotes the fact that it is for a two derivative theory of gravity.
Substituting it in equation \eqref{eq:boundaryentcur}, we get the following expression for the boundary entropy current for two derivative theory.
$$J^\mu_{(2)}=  {1\over\sqrt{g^{(b)}}} \left[\tilde t_\alpha\tilde t_\beta~\epsilon^{\alpha\alpha_1\cdots \alpha_n}\epsilon^{\beta\beta_1\cdots \beta_n} ~\chi_{\alpha_1\beta_1}\cdots \chi_{\alpha_n\beta_n}\over n!\right]^{1\over 2}  t^\mu$$
In the above expression, the vector fields $\tilde t_\mu$ and $t^\mu$ appear. They depend on our choice of mapping and naively, it seems that even in two derivative theories of gravity, the boundary entropy current might not admit a description in terms of fluid variables. But in this section, we would like to argue that this is not the case; all the factors that might not admit a derivative expansion or fluid description cancel between $t^\mu$ and $\tilde t_\mu$, and we could rewrite $J^\mu_{(2)}$ entirely in terms of fluid variables.

Note, $t_\mu=g^{(b)}_{\mu\nu}t^\nu $ and $\tilde t_\mu$ could be viewed as two vectors on the boundary with the following inner products with respect to the boundary metric
$$t_\mu\tilde t_\nu~ [g^{(b)}]^{\mu\nu} =1$$
whereas $t^\mu = e^\phi G^{\mu r}$ is a time-like vector with respect to the boundary metric\footnote{This is because at leading order in derivative expansion
$t^\mu = e^\phi G^{\mu r} = e^\phi u^\mu + {\cal O}(\partial)$. Now $u^\mu$ is a time like vector and derivative corrections can never change the sign of the leading order result}. Define the unit vector along the direction of $t^\mu$ as follows
$$\hat n^\mu \equiv{ t^\mu\over ||t||} = {n^\mu\over ||n||},~~~\text{where}~~~||t|| \equiv \sqrt{-t^\mu t^\nu g_{\mu\nu}^{(b)}}, ~~||n|| \equiv \sqrt{-n^\mu n^\nu g_{\mu\nu}^{(b)}}~~\text{and}~~ n^\mu = G^{\mu r}(r=0)$$
We can always decompose the vector $\tilde t_\mu$ in the following way
\begin{equation}\label{eq:tdecomp}
\tilde t_\mu={\hat n_\mu \over ||t||}+ V_\mu =  g^{(b)}_{\mu\nu} ~\left(\hat n^\nu \over ||t||\right)+ V_\mu,~~~\text{such that}~~V_\nu t^\nu =0
\end{equation}

Now $\chi_{\mu\nu}$ on the horizon  satisfies the following identity $t^\mu\chi_{\mu\nu}=0$. So both indices of $\chi_{\mu\nu}$ are in the directions perpendicular to $t^\mu$ or $\hat n^\mu$. It follows that in the tensor 
$$ A^{\alpha\beta}\equiv\left[~\epsilon^{\alpha\alpha_1\cdots \alpha_n}\epsilon^{\beta\beta_1\cdots \beta_n} ~\chi_{\alpha_1\beta_1}\cdots \chi_{\alpha_n\beta_n}\over n!\right]$$ 
all the indices $\{\alpha_i\}$ and $\{\beta_i\}$ in the Levi Cevita tensors are contracted 
with vectors perpendicular to $\hat n^\mu$. Hence, $A^{\alpha\beta}$ will be non-zero only when both of its free indices are projected along the direction of $\hat n$. In other words, $V_\alpha A^{\alpha\beta} = V_\beta A^{\alpha\beta} =0$, where $V_\mu$ is defined in equation \eqref{eq:tdecomp}. Therefore
\begin{equation}\label{eq:finsimp}
\begin{split}
&\tilde t_\alpha \tilde t_\beta A^{\alpha\beta} = {1\over ||t||^2}\hat n_\alpha \hat n_\beta A^{\alpha\beta}\\
\Rightarrow ~~J^\mu_{(2)} =  &{1\over\sqrt{g^{(b)}}} \left[\tilde t_\alpha\tilde t_\beta~\epsilon^{\alpha\alpha_1\cdots \alpha_n}\epsilon^{\beta\beta_1\cdots \beta_n} ~\chi_{\alpha_1\beta_1}\cdots \chi_{\alpha_n\beta_n}\over n!\right]^{1\over 2}  t^\mu \\
=  &{1\over\sqrt{g^{(b)}}} \left[\hat n_\alpha\hat n_\beta~\epsilon^{\alpha\alpha_1\cdots \alpha_n}\epsilon^{\beta\beta_1\cdots \beta_n} ~\chi_{\alpha_1\beta_1}\cdots \chi_{\alpha_n\beta_n}\over n!\right]^{1\over 2}  {t^\mu\over ||t||} \\
=  &{1\over\sqrt{g^{(b)}}} \left[\hat n_\alpha\hat n_\beta~\epsilon^{\alpha\alpha_1\cdots \alpha_n}\epsilon^{\beta\beta_1\cdots \beta_n} ~\chi_{\alpha_1\beta_1}\cdots \chi_{\alpha_n\beta_n}\over n!\right]^{1\over 2}  \hat n^\mu\\
\end{split}
\end{equation}

Note $\hat n^\mu = {n^\mu \over ||n||}$ could be entirely expressed in terms of fluid variables and the boundary metric and therefore admit derivative expansion.  
Equation \eqref{eq:finsimp} is a manifestly covariant entropy current for the boundary fluid dual to Einstein gravity, which always admits a derivative expansion.
After we know that the horizon current will translate into such a covariant `hydro-like' expression for the boundary current, we are free to choose any kind of coordinates and mapping. Even if our choice breaks all the symmetries, the final result is guaranteed to be a covariant entropy current for the dual fluid theory.

\subsection{Entropy current in higher derivative theories}\label{subsec:enthigh}
In this subsection, we would like to contrast the previous description with the scenario in higher derivative theories. 
In higher derivative theories, $j^v$ and $j^i$ have non-trivial structures constructed out of the metric functions ($X,~\omega_i,~h_{ij}$) and their derivatives in the horizon-adapted coordinates. The details of these structures will depend on the particulars of the higher derivative equations of motion. As we have seen before, once translated to boundary coordinates, the entropy current, in general, will be a vector function of both the fluid variables and the mapping variables. \\
But unlike in two derivative theories where the entropy density (as given by $\sqrt{h}$) is invariant under any reparametrization of the null generators, here affine parametrization is crucial for the construction of $j^v$ and $j^i$. This probably indicates that in a higher derivative theory, we would not be able to rearrange the formula for boundary entropy current to completely eliminate the dependence on the mapping like we have done in  Einstein gravity. So, here the key question turns out to be  whether there exists a choice of horizon to boundary map that allows us to express the final fluid entropy current entirely in terms fluid variables, without any explicit dependence on boundary coordinates (any arbitrary map, generically not compatible with derivative expansion will lead to such explicit dependence on boundary coordinates). Further,  given the non-universality of the structures appearing in $j^v$ and $j^i$ it is unlikely that we would be able the answer this question in a universal way - a single map will not work for entropy current in all higher derivative theories. However, the following simplification could be predicted on a general ground.

\begin{itemize}
\item The final fluid entropy current $J^\mu$ will not have any free `$i$' index (the spatial indices in the horizon adapted coordinates). Therefore, all the $l^\mu_i = \left(\partial x^\mu\over \partial\alpha^i\right)$ must be contracted with the inverse mapping $l^i_\mu = \left(\partial\alpha^i\over \partial x^\mu\right)$, which are the only sources of $i$ indices in $J^\mu$). Now
$$l^\mu_i l^i_\nu = \left(\partial x^\mu\over \partial\alpha^i\right)\left(\partial\alpha^i\over \partial x^\nu\right)= \delta^\mu_\nu - \left(\partial x^\mu\over \partial v\right)\left(\partial v\over \partial x^\nu\right)= \delta^\mu_\nu - t^\mu \tilde t_\nu$$
So finally, all the dependence on the mapping functions could be transferred to the dependence on $t^\mu$ and $\tilde t_\mu$.

\item $t^\mu $ could be written as $t^\mu = e^\phi n^\mu$, and it is the scalar function $\phi$ that does not admit a derivative expansion. So from the fluid point of view, the two scalar functions $\phi(x^\mu)$ and $v(x^\mu)$ could spoil the `fluid nature' of the boundary entropy current. 

\item The variations of these scalars along the direction of $n^\mu$ are constrained.
\begin{equation}\label{eq:scalconst}
(n\cdot \partial) \phi =  \left(n^\mu n^\nu\over 2\right)\left[ \partial_r \chi_{\mu\nu}\right]_{r=0},~~~~~~(t\cdot\partial)v =1~\Rightarrow~(n\cdot\partial) v = e^{-\phi}
\end{equation}
Once we `choose' these scalars on a given slice perpendicular to $n^\mu$, these equations will fix their subsequent evolution along the $n^\mu$ directions.
\item From the two equations in \eqref{eq:scalconst}, we could solve $v$ in terms of $\phi$ perturbatively using derivative expansion. This could be done as follows.\\
Define $L \equiv e^{\phi}v$. Then the equation for $L$ turns out to be
$$(n\cdot\partial) L - L(n\cdot\partial)\phi =1$$
Assume $L$ admits a derivative expansion and could be expressed entirely in terms of fluid variables, with the leading terms having zero derivatives. Since we already know that $(n\cdot\partial)\phi$ starts from zeroth order, it follows that  $(n\cdot\partial) L$ - the first term in the above equation is actually subleading in terms of derivative expansion. This allows us to solve the equation recursively generating the following infinite series
\begin{equation}\label{eq:lambda}
v \equiv e^{-\phi} L = e^{-\phi}\sum_{k=0}^\infty L_{(k)},~~\text{where}~~L_{(k)}= \left[(n\cdot\partial)L_{(k-1)}\over (n\cdot\partial)\phi\right],~~~L_{(0)} = -\left(1\over (n\cdot\partial)\phi\right)
\end{equation}
Note that this solution implies a very particular choice for the $v =0$ slice of the horizon; it is the spatial slice where  $\phi \rightarrow\infty$
Using equation \eqref{eq:lambda} we could express $\tilde t_\mu$ in terms of $\partial_\mu\phi$.
\begin{equation}\label{eq:tildet}
\begin{split}
\tilde t_\mu = e^{-\phi}\left(-L~\partial_\mu \phi + \partial_\mu L\right)
\end{split}
\end{equation}

\item $\tilde t_\mu$ must satisfy the condition $l^\mu_i \tilde t_\mu =0$ for every $i$ index (coordinates along the spatial section of the horizon)
\begin{equation}\label{eq:lmui}
\begin{split}
0=~&l^\mu_i \tilde t_\mu = e^{-\phi}\left(L~l^\mu_i\partial_\mu \phi + l^\mu_i\partial_\mu L\right)\\
\Rightarrow~&l^\mu_i\partial_\mu \phi =  { l^\mu_i\partial_\mu L\over L}
\end{split}
\end{equation}
Now we have seen that $L$ satisfies derivative expansion with the leading term being zeroth order in derivatives. So from equation \eqref{eq:lmui} we could infer that the variation of $\phi$ along the $\alpha^i$ directions also satisfies derivative expansion with the leading term being of first order. \\
%

Naively it seems that  \eqref{eq:lmui} is not consistent  because $(t\cdot\partial)$ and $(l_i\cdot\partial)$ must commute; from equation \eqref{eq:lmui}, it follows that $(t\cdot\partial)(l_i\cdot\partial)\phi $ is a second order term whereas $(l_i\cdot\partial)(t\cdot\partial)\phi $ looks like a first order term since $(t\cdot\partial)\phi$ is of zeroth order. However, we could show that the first order piece in  $(l_i\cdot\partial)(t\cdot\partial)\phi $ vanishes once we apply \eqref{eq:lmui}.

\begin{equation}\label{eq:check}
\begin{split}
&e^{-\phi}\left(l^\mu_i\partial_\mu\right)(t\cdot\partial)\phi\\
 &=e^{-\phi}\left(l^\mu_i\partial_\mu\right)(e^\phi ~n\cdot\partial)\phi \\
&=\left(l^\mu_i\partial_\mu\phi\right)(n\cdot\partial\phi)+ \left(l^\mu_i\partial_\mu\right)(n\cdot\partial\phi)\\
&=\left(l^\mu_i\partial_\mu L_{(0)}\over L_{(0)}\right)(n\cdot\partial\phi)+ \left(l^\mu_i\partial_\mu\right)(n\cdot\partial\phi) +{\cal O}\left(\partial^2\right)\\
&={\cal O}\left(\partial^2\right)
\end{split}
\end{equation}
 In the last line, we have used equation \eqref{eq:lambda}for the expression of $L_{(0)}$.
 
\item 
It turns out that the overall factors of $e^{\phi}$ finally get canceled  between $t^\mu$, $\tilde t_\mu$ and $\sqrt h$. We could see it as follows.\\
The factors of $e^\phi$ in $j^v$ or $j^i$ are determined by their boost weight. Since $j^v$ has zero boost weight,  once translated into boundary coordinates, it will not have any factor of $e^\phi$, whereas $j^i$  having boost weight one, will carry a single factor of $e^\phi$. We have already seen $\sqrt h$, expressed in terms of boundary coordinates, carries a factor of $e^{-\phi}$ from the $||t||$ factor in the denominator (see equation \eqref{eq:finsimp}). Hence in the expression $\sqrt{h}\left( j^v t^\mu + l^\mu_i j^i\right)$ all factors of overall $e^\phi$ cancel.

Therefore, once we fix $v$ in terms of $\phi$ using equation \eqref{eq:lambda},  the `non-fluid' function that still remains in our construction is the derivative of $\phi$ along the directions perpendicular to $n^\mu$ .\\
\end{itemize}

\section{Entropy current in Einstein-Gauss-Bonnet theory}\label{sec:entcurHOR}
In this section, we shall specialize to Einstein-Gauss-Bonnet theory. The entropy  density and the entropy current for black holes in Einstein-Gauss-Bonnet theory have been worked out in \cite{our1912,our2105}. Using the horizon to boundary map, we shall rewrite the current in boundary coordinates. At this stage, we shall not use any derivative or amplitude expansion. We shall see that the final expressions will explicitly depend on the `non-fluid' variables through $\tilde t_\mu$ and $\partial_\mu\phi$.\\
Note that any term or factor that could be expressed as a product of metric components in boundary coordinates and their boundary derivatives are fluid variables.  For example, the Christoffel symbols with respect to the bulk metric in boundary coordinates are always fluid variables.

In the end, we shall substitute the details of the bulk metric in Gauss-Bonnet theory dual to hydrodynamics. Since the metric is known up to the first order in derivative expansion, the boundary entropy current thus generated will also be correct only up to the first order. As mentioned before, up to this order the entropy current will turn out to be trivial; it is simply equal to  what it was for Einstein gravity. All the new terms generated by Gauss-Bonnet Action contribute to the boundary entropy current only in second order.

\subsection{ $j^v$ and $j^i$ in terms of `fluid' and `non fluid'  data}
We shall first quote the expression for entropy density and the spatial entropy current for black holes in Gauss-Bonnet theory as given in \cite{our1912,our2105}.

The final form of the entropy density and spatial entropy current density particular to Gauss-Bonnet theory is given as follows.

\begin{equation}\label{eq:entcurhor}
\begin{split}
&j^v={\sqrt h}\left(1 + 2\alpha^2{\cal R}\right),~~~~
j^i= -4\alpha^2\left(\nabla_jK^{ij}-\nabla^i K \right)
\end{split}
\end{equation}
where
\begin{equation}\label{notation}
\begin{split}
 &h = \text{determinant of $h_{ij}$}\\
 &{\cal R}
 =\text{ intrinsic curvature evaluated w.r.t the  $h_{ij}$}\\
 &\nabla_i = \text{covariant derivative with respect to $h_{ij}$}\\
& K_{ij}\equiv {1\over 2} \partial_v h_{ij},~~~K \equiv h^{ij}K_{ij}\\
&\text{Lowering or raising of indices are done w.r.t $h_{ij}$ with $h^{ij}$ being the inverse}
\end{split}
\end{equation}
One key simplifying factor here is that neither $j^v$ nor $j^i$ needs any information about how the horizon data changes as one moves away from the horizon or, more precisely, the $r$ derivatives of the metric functions. This, in turn, implies that to evaluate the current, we need only the leading coefficients in the coordinate transformation as described in equation \eqref{coordi-1}.
In the previous subsection, we have already determined the expression for $\sqrt{h}$ in terms of fluid data. In this subsection, we shall compute $\nabla_k K_{ij}$ with appropriate index contractions for $j^il^\mu_i$ and ${\cal R}$ for $j^v$
 

\subsubsection{Extrinsic curvature and its covariant derivatives}
The extrinsic curvature is defined as $K_{ij} ={1\over 2}\partial_v h_{ij}\vert_{r=0}$. On the horizon, the $r=0$ hypersurface,  $h_{ij}$ is simply related to $\chi_{\mu\nu}$.
\begin{equation}\label{eq:hchi}
\begin{split}
h_{ij} =l^\mu_i l^\nu_j\chi_{\mu\nu}
\end{split}
\end{equation}
Here we have used the fact that $\left(\partial\rho\over\partial \alpha^i\right) $ vanishes on the horizon. Now, using the fact that $\partial_v = t\cdot\partial$, we could determine $K_{ij}$ as
\begin{equation}\label{eq:KIJ}
\begin{split}
&K_{ij} = l^\mu_i l^\nu_j {\cal K}_{\mu\nu}~~~ 
\text{where}~~~{\cal K}_{\mu\nu} = -t^\alpha\Gamma_{\alpha,\mu\nu}\end{split}
\end{equation}
Here we have used the fact that
$$(t\cdot\partial)l^\mu_i = (l_i\cdot\partial)t^\mu,~~~\text{and}~~~\chi_{\mu\nu}(l_i\cdot\partial)t^\mu=-t^\mu(l^i\cdot\partial)\chi_{\mu\nu}$$
Now we have to compute its covariant derivative. The following structure would prove useful for our computation. Note, for any boundary  tensor with lower $\{\mu,\nu\} $ indices, we could define the following horizon tensor with $\{i,j\}$ indices
$$T_{i_1i_2\cdots i_n} =l^{\mu_1}_{i_1}l^{\mu_2}_{i_2}\cdots l^{\mu_n}_{i_n}~ {\cal T}_{\mu_1\mu_2\cdots \mu_n}$$
Now it turns out that the covariant derivative of the above tensor $\nabla_j T_{i_1i_2\cdots i_n}$ 
also has a similar expression in terms of $\{\mu,\nu\}$ indices of the boundary coordinates. We could write it in the following way
\begin{equation}\label{eq:covderi}
\begin{split}
\nabla_j T_{i_1i_2\cdots i_n}= l^\nu_jl^{\mu_1}_{i_1}l^{\mu_2}_{i_2}\cdots l^{\mu_n}_{i_n}~ \left[{\cal D}_\nu{\cal T}_{\mu_1\mu_2\cdots \mu_n}\right]
\end{split}
\end{equation}
where ${\cal D}_\nu$ is a new covariant derivative with its connection defined as
\begin{equation}\label{eq:newconn}
\begin{split}
&\tilde\Gamma^\nu_{\alpha\beta} = \bar \chi^{\nu\theta}\Gamma_{\theta, \alpha\beta} + t^\nu\partial_\alpha\tilde t_\beta\\
\text{where}~& \tilde t_\mu\equiv \left(\partial v\over\partial x^\mu\right),~~\bar\chi^{\mu\nu} = {\Delta^\mu}_\alpha{\Delta^\nu}_\beta~ \chi^{\alpha\beta},~~~{\Delta^{\alpha}}_\beta \equiv \delta^\alpha_\beta - t^\alpha\tilde t_\beta
\end{split}
\end{equation}
One could easily show these structures by acting the covariant derivatives on vectors and recursively using the relations for higher indexed tensors.
Note the new connection $\tilde \Gamma^\mu_{\alpha\beta}$ is also symmetric in its lower two indices. The other mixed tensor we defined here is actually a projector to constant $v$ slices of the horizon because 
$$t^\alpha {\Delta^\beta}_\alpha = {\Delta^\alpha}_\beta~\tilde t_\alpha =0$$
Using these structures, we could see that 
\begin{equation}\label{eq:covderi2}
\begin{split}
\nabla_k K_{ij} =~& l^\alpha_kl^\mu_i l^\nu_j {\cal D}_\alpha{\cal K}_{\mu\nu}\\
=~& l^\alpha_kl^\mu_i l^\nu_j~ \left[\partial_\alpha{\cal K}_{\mu\nu} - \tilde\Gamma^\beta_{~\alpha\mu} {\cal K}_{\beta\nu} - \tilde\Gamma^\beta_{~\alpha\nu} {\cal K}_{\mu\beta}\right]\\
\text{where}~~&{\cal K}_{\mu\nu} = -t^\alpha\Gamma_{\alpha,\mu\nu}
\end{split}
\end{equation}
The spatial current on the horizon will add the following contribution to the boundary entropy current
 $$J_{space}^\mu =\frac{1}{\sqrt{g^{(b)}}} \Delta~l^\mu_a \left(\sqrt{h}~j^a \right)=-4\alpha^2~\frac{1}{\sqrt{g^{(b)}}}\Delta~l^\mu_a \sqrt{h}\left(h^{ki}h^{ja} - h^{ka} h^{ij}\right)\left(\nabla_k K_{ij}\right)$$

Now using the identity $h^{ij} l^\mu_i l^\nu_j = \bar \chi^{\mu\nu}$, we finally get the following expression for the space part of the entropy current
\begin{equation}\label{entspace}
\begin{split}
J_{space}^\mu= -4\alpha^2\frac{1}{\sqrt{g^{(b)}}}\sqrt{H}\left(\bar\chi^{\gamma\alpha}\bar\chi^{\mu\beta} - \bar\chi^{\gamma\mu}\bar\chi^{\alpha\beta}\right)\left({\cal D}_\gamma{\cal K}_{\alpha\beta}\right)~~\text{where}~~ H \equiv \frac{\hat n_\mu\hat n_\nu \epsilon^{\mu\mu_1\cdots \mu_n}\epsilon^{\nu\nu_1\cdots \nu_n}\chi_{\mu_1\nu_1}\cdots \chi_{\mu_n\nu_n}}{n!}
\end{split}
\end{equation}

\subsubsection{Intrinsic Ricci scalar}

For the temporal part of the entropy current, we need to compute the intrinsic Ricci scalar of the constant $v$ slices of the horizon. \\
In this section, we note down the calculation for the Ricci scalar, $\mathcal{R}$, with respect to $h_{ij}$.\\\\

We start with the expression for Riemann tensor
\begin{equation}
\begin{split}
\mathcal{R}^a_{bcd}=\partial_c\Gamma^a_{bd}+\Gamma^a_{cm}\Gamma^m_{bd}-(c\leftrightarrow d)
\end{split}
\end{equation}
Now we will process $\partial_c\Gamma^a_{bd}$ in the following way
\begin{equation}
\begin{split}
\partial_c\Gamma^a_{bd}&=\partial_c\left(h^{ap}\Gamma_{p,bd}\right)\\
&=\partial_ch^{ap}\Gamma_{p,bd}+h^{ap}\partial_c\Gamma_{p,bd}\\
&=-h^{aq}\Gamma^p_{cq}\Gamma_{p,bd}-\Gamma^a_{cr}\Gamma^r_{bd}+h^{ap}\partial_c\Gamma_{p,bd}
\end{split}
\end{equation}
where in the last line, we have used
\begin{equation}
\begin{split}
&\partial_ch^{ap}=-h^{aq}h^{pr}\partial_ch_{rq}\\
&\partial_ch_{rq}=\Gamma_{r,cq}+\Gamma_{q,cr}
\end{split}
\end{equation}
Hence, we have
\begin{equation}
\begin{split}
\partial_c\Gamma^a_{bd}+\Gamma^a_{cr}\Gamma^r_{bd}=-h^{aq}\Gamma^p_{cq}\Gamma_{p,bd}+h^{ap}\partial_c\Gamma_{p,bd}
\end{split}
\end{equation}
So, we can write the expression for the Riemann tensor in the following form
\begin{equation}
\begin{split}
\mathcal{R}_{abcd}=\partial_c\Gamma_{a,bd}-\Gamma^p_{ca}\Gamma_{p,bd}-\left(\partial_d\Gamma_{a,bc}-\Gamma^p_{da}\Gamma_{p,bc}\right)
\end{split}
\end{equation}
Now the expression for the $\Gamma_{k,ij}$ in the following
\begin{equation}
\begin{split}
\Gamma_{k,ij}=l^{\mu}_il^{\nu}_jl^{\alpha}_k\Gamma_{\alpha,\mu\nu}+\chi_{\mu\nu}~l^{\mu}_k\left(l_i\cdot \partial l^{\nu}_j\right)
\end{split}
\end{equation}
Then we can process $\partial_c\Gamma_{a,bd}$ in the following way
\begin{equation}
\begin{split}
\partial_c\Gamma_{a,bd}=\left(l_c\cdot\partial\right)\left[l^{\mu}_bl^{\nu}_dl^{\alpha}_a\Gamma_{\alpha,\mu\nu}+\chi_{\mu\nu}~l^{\mu}_a\left(l_b\cdot\partial l^{\nu}_d\right)\right]
\end{split}
\end{equation}
And also,
\begin{equation}
\begin{split}
&\Gamma^p_{ca}\Gamma_{p,bd}\\
&=h^{pm}\Gamma_{m,ca}\Gamma_{p,bd}\\
&=\bar{\chi}^{\alpha_1 \alpha}l^{\mu}_cl^{\nu}_al^{\mu_1}_bl^{\nu_1}_d\Gamma_{\alpha,\mu\nu}\Gamma_{\alpha_1,\mu_1\nu_1}+\Delta^{\alpha}
_{\nu_1}l^{\mu}_cl^{\nu}_a\Gamma_{\alpha,\mu\nu}\left(l_b\cdot\partial l^{\nu_1}_d\right)+\Delta^{\alpha_1}_{\nu}l^{\mu_1}_bl^{\nu_1}_d\left(l_c\cdot\partial l^{\nu}_a\right)\Gamma_{\alpha_1,\mu_1 \nu_1}\\
&~~+\chi_{\nu\nu_1}\left(l_c\cdot\partial l^{\nu}_a\right)\left(l_b\cdot\partial l^{\nu_1}_d\right)
\end{split}
\end{equation}
where, we have defined $\bar\chi^{\alpha\beta}=l^{\alpha}_il^{\beta}_jh^{ij}$ and used the fact that $\bar\chi^{\alpha\beta}\chi_{\beta\nu}=\Delta^{\alpha}_{\nu}$. \\\\
Then we have
\begin{equation}
\begin{split}
&\partial_c\Gamma_{a,bd}-\Gamma^p_{ca}\Gamma_{p,bd}\\
=&\Gamma_{\alpha,\mu\nu}\left[l^{\nu}_dl^{\alpha}_a\left(l_c\cdot\partial\right)l^{\mu}_b+l^{\mu}_bl^{\alpha}_a\left(l_c\cdot\partial\right)l^{\nu}_d+l^{\mu}_cl^{\alpha}_a\left(l_b\cdot\partial\right)l^{\nu}_d\right]\\
&+l^{\mu}_bl^{\nu}_dl^{\alpha}_al^{\beta}_c\left[\partial_{\beta}\Gamma_{\alpha,\mu\nu}-\bar\chi^{\alpha_1\alpha_2}\Gamma_{\alpha_2,\beta\alpha}\Gamma_{\alpha_1,\mu\nu}\right]+\chi_{\mu\nu}l^{\mu}_a\left[\left(l_c\cdot\partial\right)\left(l_b\cdot\partial l^{\nu}_d\right)\right]\\
&+t^{\alpha}\Gamma_{\alpha,\mu\nu}\left[l^{\mu}_cl^{\nu}_a\tilde{t}_{\nu_1}\left(l_b\cdot \partial l^{\nu_1}_d\right)+l^{\mu}_bl^{\nu}_d\tilde{t}_{\nu_1}\left(l_c\cdot\partial l^{\nu_1}_a\right)\right]
\end{split}
\end{equation}
Hence, we have the expression for $\mathcal{R}_{abcd}$ as
\begin{equation}
\begin{split}
\mathcal{R}_{abcd}=&\left[\partial_{\beta}\Gamma_{\alpha,\mu\nu}-\bar\chi^{\alpha_1\alpha_2}\Gamma_{\alpha_2,\beta\alpha}\Gamma_{\alpha_1,\mu\nu}\right]\left[l^{\mu}_bl^{\nu}_d l^{\alpha}_{a}l^{\beta}_c-l^{\mu}_bl^{\nu}_cl^{\alpha}_al^{\beta}_d\right]\\
&+t^{\alpha}\Gamma_{\alpha,\mu\nu}\partial_{\delta}\tilde{t}_{\nu_1}\left[l^{\mu}_{b}l^{\nu}_cl^{\delta}_dl^{\nu_1}_a+l^{\mu}_dl^{\nu}_al^{\delta}_bl^{\nu_1}_c-l^{\mu}_bl^{\nu}_dl^{\delta}_cl^{\nu_1}_a-l^{\mu}_cl^{\nu}_al^{\delta}_bl^{\nu_1}_d\right]
\end{split}
\end{equation}
So, finally we have
\begin{equation}
\begin{split}
\mathcal{R}=&h^{ac}h^{bd}\mathcal{R}_{abcd}\\
=&\left[\bar\chi ^{\mu\nu}\bar\chi^{\alpha\beta}-\bar\chi^{\alpha\nu}\bar\chi^{\mu\beta}\right]\left[\partial_{\beta}\Gamma_{\alpha,\mu\nu}-\bar\chi^{\alpha_1\alpha_2}\Gamma_{\alpha_2,\beta\alpha}\Gamma_{\alpha_1,\mu\nu}-2t^{\alpha_1}\Gamma_{\alpha_1,\mu\nu}\partial_{\alpha}\tilde{t}_{\beta}\right]
\end{split}
\end{equation}\\
Hence, we finally get the following expression for the intrinsic Ricci scalar
\begin{equation}\label{eq:Ricciscalar}
\begin{split}
{\cal R} = \left(\bar\chi^{\mu_1\nu_1}\bar\chi^{\mu_2\nu_2}-\bar\chi^{\mu_1\nu_2}\bar\chi^{\mu_2\nu_1}\right)\bigg[\partial_{\mu_1}\Gamma_{\nu_1,\mu_2\nu_2} - \bar\chi^{\alpha_1\alpha_2}\Gamma_{\alpha_1,\mu_1\nu_1}\Gamma_{\alpha_2,\mu_2\nu_2}-2 t^\alpha\Gamma_{\alpha,\mu_1\nu_1}\left(\partial_{\mu_2}\tilde t_{\nu_2}\right)\bigg]
\end{split}
\end{equation}
\subsubsection{Separating `fluid' and `non-fluid' terms}
The final form of the entropy current written in terms of boundary coordinates $\{x^\mu\}$ is
\begin{equation}\label{eq:entcurfin}
\begin{split}
&J^\mu = J^\mu_{space} + J^\mu_{time}\\
\text{where}~~&J^\mu_{space} = -4\alpha^2\frac{1}{\sqrt{g^{(b)}}}\frac{\sqrt{H}}{||t||}\left(\bar\chi^{\gamma\alpha}\bar\chi^{\mu\beta} - \bar\chi^{\gamma\mu}\bar\chi^{\alpha\beta}\right)\left({\cal D}_\gamma{\cal K}_{\alpha\beta}\right)\\
&J^\mu_{time}=\frac{1}{\sqrt{g^{(b)}}}\sqrt{H} \left(1+2\alpha^2 {\cal R}\right) \hat n^\mu\\
\end{split}
\end{equation}

${\cal R}$ is given in equation \eqref{eq:Ricciscalar}.\\
In this expression of the current, most of the terms are `fluid' terms in the sense that they depend solely on the metric components and their derivatives written in boundary coordinates. The exceptions are those terms where one has explicit $\tilde t_\mu$, e.g., in $\Delta^\mu_\nu = \delta^\mu_\nu - t^\mu\tilde t_\nu$.
These terms could be further processed by expressing $\tilde t_\mu$ in terms of $\partial_\mu\phi$ using equation\eqref{eq:tildet}. The expressions turn out to be too  big to be presented here. We have collected them in appendix (\ref{app:decomp}).

In the final stage, we would like to evaluate this current on the hydrodynamic metric correctly up to first order in derivative expansion. However, just looking at equation \eqref{eq:entcurfin}, we could figure out that $J^\mu_{space}$ is of second order. This is because $\Gamma_{\alpha,\mu\nu}$ is always of first order in terms of derivative expansion and so is ${\cal K}_{\alpha\beta}\sim t^\mu\Gamma_{\mu,\alpha\beta}$. It follows that $J^\mu_{space}\sim {\cal D}_\gamma{\cal K}_{\alpha\beta}\sim {\cal O}(\partial^2)$. Using a similar argument, we could show that ${\cal R}$ is also of ${\cal O}(\partial^2)$, where we have used the fact that $\tilde t_\mu$ is of order ${\cal O}(1)$ in terms of derivative expansion. Therefore, up to first order in derivative expansion, there will not be any contribution to the entropy current from the  Gauss-Bonnet correction. To have any non-trivial result, we need to go at least one higher order in derivative expansion, which we leave for future work.


\section{Future Directions}\label{sec:future}
If we follow our construction, the boundary entropy current will involve one `non-fluid'  function, the scalar field $\phi$, whose exponential relates the two different parametrizations of the horizon null generator. But the fluid entropy current must not have any other  field other than the fluid velocity  and its local temperature. So the next natural question is whether we could use the non-uniqueness of the currents on both horizon and the boundary side to remove this unwanted $\phi$ dependence. Einstein-Gauss-Bonnet theory is the simplest well-studied example where such currents and ambiguities could be explicitly constructed and tested. But unfortunately to the order that we have worked here, no such fixing is required since up to this order all the non-trivial structures that have this `non fluid' $\partial_\mu\phi$ factor just vanish.
So our future goal would be to extend this calculation to order ${\cal O}(\partial^2)$. In this section we shall set up the stage for this future calculation.

\subsection{Conditions of stationarity}\label{sec:stationarity}
As mentioned before, the entropy current and the entropy density in higher derivative theories work only for horizons where the amplitude (let's denote it as $\epsilon$) of the dynamics is small and could be treated perturbatively. Moreover, the construction in \cite{our1912,our2105} works only up to the linear order in $\epsilon$.  So we should not expect the dual fluid entropy current to do any better. In other words, while applying formula \eqref{eq:boundaryentcur}, we should ignore all terms that are of ${\cal O}(\epsilon^2)$ or higher in $\chi_{\mu\nu}$, $t^\mu$ or $\tilde t_\mu$. \\
Now derivative expansion is not the same as amplitude expansion. It is quite possible to have terms that are linear in $\epsilon$ but higher order in terms of derivative. So we need to have a clean prescription to identify fluid data that are linear in amplitude (but in principle, could have multiple derivatives). 

A stationary fluid on the boundary (where both the boundary metric and the fluid configuration admit at least one Killing vector) should be dual to a stationary bulk metric with a Killing horizon.  In other words, 
the Killing vector on the boundary could be extended to a bulk Killing vector, which on the horizon reduces to the Killing generator of the horizon. In terms of equations, what we mean is the following.
Suppose $\xi = \xi^A\partial_A$ is the bulk Killing vector. \\
Since it reduces to the generator of the horizon (the $r=0$ hypersurface in our choice of coordinates) 
$$\lim_{r\rightarrow 0} \xi^r =0,~~~\lim_{r\rightarrow 0}\xi^\mu\propto G^{\mu r}\vert_{r=0}$$
Further, $\xi^A\partial_A$ should reduce to the boundary Killing vector $\xi^\mu_{(b)}\partial_\mu$ in the limit $r\rightarrow\infty$
$$\lim_{r\rightarrow\infty}\xi^r =0,~~~\lim_{r\rightarrow\infty}\xi^\mu = \xi^\mu_{(b)}$$
Now for our analysis, we shall assume\footnote{This assumption could be justified as follows. Let's choose a coordinate system where $\xi^A\partial_A = \partial_\tau$, i.e. $\tau$ is the parameter along the integral curve of $\xi^A$. The Killing coordinate is $\tau$, and hence, the metric could be expressed such that all of its components are independent of $\tau$. Since the boundary metric is just the boundary limit of the bulk metric, its components should also be independent of $\tau$. The same should be true of the fluid variables like velocity and temperature as the bulk metric components are functions of these variables only. Therefore the same $\tau$ will also be a Killing coordinate from the perspective of the stationary  boundary fluid.} that~~~~
$$\lim_{r\rightarrow 0}\xi^\mu =G^{\mu r}\vert_{r=0} = \xi^\mu_{(b)}$$
The above condition will result in a set of constraints both on the fluid data and the horizon data (vanishing of some particular fluid/ horizon  structures), respectively.  Any violation of these constraints will be a departure from stationarity and, therefore, generically of order ${\cal O}(\epsilon)$ terms.
We have a clean classification of such terms on the horizon side and using the map, we could translate them to the fluid side. The ${\cal O}(\epsilon)$ terms, thus derived on the fluid side, should be automatically compatible with constraints of stationarity (and departure from it) as expected from any stationary fluid configuration.

Product of two such order ${\cal O}(\epsilon)$ terms will be order ${\cal O}(\epsilon)^2$ and therefore neglected.

\subsection{Choice of Fluid Frames}\label{subsec:fluidframe}
In section \ref{sec:coordinates}, we have presented the metric dual to boundary fluid dynamics (see equation \eqref{fluidmetric}). This metric is written in terms of fluid velocity ($u^\mu$) and temperature ($T$). But as one goes to higher order in derivative expansion, one has the freedom to redefine the velocity and the temperature of the fluid. This ambiguity is present in fluid dynamics itself and is usually fixed by a specific choice of fluid frames. Now fluid dynamics is about the dynamics of the stress tensor and other conserved charges of the system. So the fluid frames are also usually defined in terms of the stress tensor or the currents. For example, in `Landau frame' the velocity of the energy flow is defined as $u^\mu$. This implies that $u^\mu$ is the unique time-like eigenvector of the stress tensor (normalized). Once $u^\mu$  (and temperature) is unambiguously defined, the dual bulk metric is constructed. A given definition of the fluid frame amounts to a given boundary condition for the metric function while solving for the bulk metric.

In this section, we shall adopt a different choice of fluid frame which would be more suitable for our purpose, and in particular, for the description of equilibrium. 
We shall define our new velocity $u^\mu$ as 
$$u^\mu \equiv \hat n^\mu,~~\text{where}~~n^\mu = G^{\mu r}\vert_{r=0}~~\text{and}~~\hat n^\mu = {n^\mu\over \sqrt{-n^\mu n^\nu g^{(b)}_{\mu\nu}}}$$
For brevity, we shall denote this choice of velocity as `Gravity frame'. One could choose this frame only if the fluid admits a gravity dual.
Note that 
$$u^\mu\vert_\text{gravity frame} = u^\mu\vert_\text{Landau frame} +{\cal O}\left(\partial\right)$$
So in zeroth order in derivative, these two definitions of velocity agree as they should. In fact, it turns out that even at first order in derivative expansion, these two velocities agree; the difference starts only at second order. However, since in this note, our computations are correct only up to first order in derivative expansion, this frame redefinition becomes particularly simple for us. Basically, it says there is no transformation at all up to first order in derivatives.
 
 \subsection{Metric Dual to Hydrodynamics in Gauss-Bonnet Theory in Gravity frame}\label{sec:metFLUID}
The metric dual to hydrodynamic in Einstein-Gauss-Bonnet theory has been worked out in \cite{Dutta_2008} up to first order in derivative expansion. However, in \cite{Dutta_2008} the main concern was boundary hydrodynamics and therefore, the author has worked in a slightly different gauge than what is described in equation \eqref{shift}. In this subsection, we shall work out the same metric, but in the gauge most convenient for our purpose , i.e., using the gravity frame described in the previous section.

The action for the full "Einstein + Gauss-Bonnet" theory is given by \footnote{Here, we have used the convention $4 G_5 = 1$ (where $G_5$ is the Newton's constant in five dimensions) to have only the horizon area term without any extra proportionality constants as the entropy of the Einstein theory. Accordingly, the proportionality constant in $S_E$ and $S_{GB}$ have been modified from those used in \cite{Dutta_2008}.}
\begin{equation}
    \begin{split}
        S &= S_E + \alpha^2~ S_{GB} \\
        S_E &={\color{black}-\frac{1}{4\pi}} \int d^5 x~ \sqrt{-g}  (R - 2 \Lambda) \\
        S_{GB} &={\color{black}-\frac{1}{4\pi}} \int d^5 x~ \sqrt{-g}  ( R^2 - 4 R^{AB} R_{AB} + R^{ABCD} R_{ABCD} )
    \end{split}
\end{equation}
We will parametrize $\Lambda$ \footnote{In \cite{Dutta_2008}, to ensure the fact that the boundary metric is exactly equal to the Minkowski metric - $\eta_{\mu\nu}$, the author has to scale the boundary coordinates in an $\alpha$ dependent manner. As a result in the final covariant bulk metric the component $G_{\rho\mu}$ is no longer equal to $-u_\mu$, rather just proportional to it with an $\alpha$ dependent constant as proportionality factor.  However, in our analysis we have crucially used the fact that $G_{\rho\mu} = -u_\mu$ and also the calculation simplifies if the boundary metric is just equal to $\eta_{\mu\nu}$. It turns out if we want to impose both these conditions on the bulk metric, we need to scale the consmological constant. } as 
$\Lambda=-6\left(1-2\alpha^2\right)$.\\
The equations of motion of the full theory are given by
\begin{equation}
    \begin{split}
        E_{MN} &= \left( R_{MN} - \frac{1}{2} g_{MN} R + \Lambda g_{MN}  -  \frac{1}{2} \alpha^2 ~g_{MN}  ( R^2 - 4 R^{AB} R_{AB} + R^{ABCD} R_{ABCD} ) \right) \\
        &+ \alpha^2  \left(   4 R_{MPQL} R_N^{~PQL} - 4 R^{PQ} R_{MPNQ} - 4R_M^{~P} R_{NQ} + 2 R~R_{MN} \right)
    \end{split}
\end{equation}

The black-brane metric which is dual to a boundary fluid and solves these equations of motion up to first order in derivatives as well as in $\alpha^2$ is given by
\begin{equation}
    ds^2 = -2   u_\mu dx^\mu dr + \chi_{\mu\nu} dx^\mu dx^\nu
\end{equation}

Note that in this gauge, the boundary metric will be of the form $g^{(b)}_{\mu\nu}=\eta_{\mu\nu}$ and lowering and raising of the boundary indices have to be done w.r.t $g^{(b)}_{\mu\nu}$. $\chi_{\mu\nu}$ can be expressed as
\begin{equation}\label{gbmetric}
\begin{split}
\chi_{\mu\nu}=&-r_H^2~f\left(\frac{r}{r_H}\right)~u_{\mu}u_{\nu}+r_H^2~K\left(\frac{r}{r_H}\right)~P_{\mu\nu}+r_H~F\left(\frac{r}{r_H}\right)~\sigma_{\mu\nu}+r_H~V\left(\frac{r}{r_H}\right)\left(u_{\mu}a_{\nu}+u_{\nu}a_{\mu}\right)\\
&+\theta\left(r_H~S_1\left(\frac{r}{r_H}\right)~u_{\mu}u_{\nu}+r_H~S_2\left(\frac{r}{r_H}\right)~P_{\mu\nu}\right)
\end{split}
\end{equation}
As mentioned before, here, $r_H$ is the scale associated with the black hole solution. The functions used in \eqref{gbmetric} are defined as
\begin{equation}\label{gbmetricfunctions}
\begin{split}
f(x)=&\left(1+x\right)^2\left[1-\left(\frac{1}{1+x}\right)^4\right]-2~\alpha^2\frac{\left[\left(1+x\right)^4-1\right]}{\left(1+x\right)^6}\\
K(x)=&\left(1+x\right)^2\\
V(x)=&-x\\
S_1(x)=&\frac{2x}{3}\\
S_2(x)=&0\\
F(x)=&F_0(x)+\alpha^2~F_{\alpha}(x)\\
\end{split}
\end{equation}
with,
\begin{equation}
\begin{split}
F_0(x)=&\frac{1}{2} (x+1)^2 \left(-4 \log (x+1)+2 \log (x+2)+\log (x (x+2)+2)-2 \tan ^{-1}(x+1)+\pi
   \right)\\
 F_{\alpha}(x)=&\frac{1}{(1+x)^2}\bigg[ \pi  (x+1)^4-4 x (x (x+3)+3)-4 (x+1)^4 \log (x+1)+4 \log (x+2)+3 \log (x (x+2)+2)\\
 &+x (x+2) (x
   (x+2)+2) (4 \log (x+2)+3 \log (x (x+2)+2)-2 \log ((x+2) (x (x+2)+2)))\\
   &-2 \log ((x+2) (x
   (x+2)+2))-2 (x+1)^4 \tan ^{-1}(x+1)-1\bigg]
\end{split}
\end{equation}
and the fluid variables $\theta$ and $\sigma_{\mu\nu}$ and the projector $P_{\mu\nu}$  are given by
\begin{equation}
    \begin{split}
    P_{\mu\nu}&=g^{(b)}_{\mu\nu}+u_{\mu}u_{\nu}\\
        \theta &= \partial \cdot u \\
        \sigma_{\mu\nu} &= P_\mu^\alpha P_\nu^\beta \partial_{(\alpha} u_{\beta)}
    \end{split}
\end{equation}
 
 
\subsection{Stationary solution in Gravity frame}\label{subsec:gravityframe}
 In a stationary metric with horizon located at $r=0$, the Killing vector is $\xi^\mu \propto G^{\mu r}\vert_{r=0}$. According to our assumption
 $$\xi^\mu_{(b)} \propto G^{\mu r}\vert_{r=0},~~\Rightarrow~~\xi^\mu_{(b)}\propto u^\mu~~~\text{in Gravity frame}$$
 Now in a stationary situation $G^{\mu r}$ is proportional to the Killing vector, both for the Bulk and the boundary metric. Therefore, in case of stationary fluid, this particular choice of frame amounts to choosing the fluid velocity in the direction of the Killing vector for the boundary metric.
 
 In this subsection, we shall start from the assumption that $\xi^A\partial_A = \xi^r\partial_r + F(r,x^\mu) u^\mu\partial_\mu$. Then we shall derive the conditions $u^\mu$ must satisfy so that $\xi^A\partial_A$ is a bulk Killing vector. We shall see that  $u^\mu$ will turn out to be proportional to the boundary Killing vector as expected, with its shear tensor and expansion vanishing everywhere.\\
Now we will show that if we have a Killing vector proportional to the fluid velocity $u^{\mu}$, then the expansion and shear tensor will vanish. We will also get  constraints on the proportionality constant such that this condition is satisfied.\\
We will start by writing the fluid metric in a way such that the horizon is located at the origin of the radial coordinate.
\begin{equation}
\begin{split}
ds^2=-2u_{\mu}~dx^{\mu}~dr-r_H^2~f\left(r/r_H\right)\left(u_{\mu}dx^{\mu}\right)^2+(r+r_H)^2P_{\mu\nu}~dx^{\mu}~dx^{\nu}+\chi_{\mu\nu}^{(1)}~dx^{\mu}~dx^{\nu}
\end{split}
\end{equation}
where, $\chi^{(1)}_{\mu\nu}$ contains terms first order in derivative of the fluid variables.\\
%
Then the killing vector will have the following form
\begin{equation}
\begin{split}
\xi^A\partial_A\propto G^{\mu r}\vert_{r=0}=F~u^{\mu}\partial_{\mu}
\end{split}
\end{equation}
where, $F$ is the proportionality constant.\\
In covariant form this becomes
\begin{equation}
\begin{split}
\xi_Adx^A=F~dr+F\left[r_H^2~f\left(r/r_H\right)~u_{\alpha}+u^{\mu}\chi^{(1)}_{\mu\alpha}\right]~dx^{\alpha}
\end{split}
\end{equation}
Now we will solve for the Killing equation on this and write down the conditions it will give on $F$ and $\chi^{(1)}$.\\
The Killing equation is 
\begin{equation}
\begin{split}
\nabla_A\xi_B+\nabla_B\xi_A=0
\end{split}
\end{equation}
The $\left(r,r\right)$ component of which will give the following condition
\begin{equation}\label{drF}
\begin{split}
\partial_rF=0
\end{split}
\end{equation}
The $\left(r,\mu\right)$ component will give
\begin{equation}\label{xi_r_mu1}
\begin{split}
\partial_{\mu}F-F~a_{\mu}=0
\end{split}
\end{equation}
where, $a_{\mu}=\left(u\cdot\partial\right)u_{\mu}$.\\
The $\left(\mu,\nu\right)$ component will give
\begin{equation}
\begin{split}
&r_H^2~f\left(r/r_H\right)~\bigg[u_{\mu}\left(\partial_{\nu}F-F~a_{\nu}\right)+u_{\nu}\left(\partial_{\mu}F-F~a_{\mu}\right)\bigg]+2F(r+r_H)^2\sigma_{\mu\nu}\\
&+F~r(r+r_H)\frac{2\theta}{D-2}P_{\mu\nu}+r_H~F\left[2r_H~f\left(r/r_H\right)-r~f^{\prime}\left(r/r_H\right)\right]\frac{\theta}{D-2}u_{\mu}u_{\nu}=0
\end{split}
\end{equation}
where we have used the following identity and fluid constraint equation,
$\partial_{\mu}u_{\nu}=\sigma_{\mu\nu}+\omega_{\mu\nu}-u_{\mu}a_{\nu}+\frac{\theta}{D-2}P_{\mu\nu}$ and $\frac{(u\cdot\partial)r_H}{r_H}+\frac{\theta}{D-2}=0$.\\

Now to be consistent with  \eqref{xi_r_mu1} we should have
\begin{equation}
\begin{split}
&\theta=0,~~~
\sigma_{\mu\nu}=0
\end{split}
\end{equation} 
Hence, we could show that with vanishing shear tensor and expansion, $F~u^{\mu}$ is actually a Killing vector with $F$ satisfying \eqref{drF} and \eqref{xi_r_mu1}. \\
Note that $F=\frac{1}{r_H}$ is a solution to \eqref{drF} and \eqref{xi_r_mu1}. Also note that in \cite{Caldarelli_2009} the Killing vector $\xi^{\alpha}={c\over T }~u^\alpha$ where, $T$ is the local temperature $T=\left(\frac{D-1}{4\pi}\right)r_H$ and $c$ is a constant. Hence upto an overall constant the two Killing vectors are equivalent.\\
Hence, these stationarity conditions are identical to the ones derived in  \cite{Caldarelli_2009}  from the perspective of a stationary  boundary fluid.


\section{Conclusion}\label{sec:conclude}

The construction of \cite{our1912,our2105} gives  an expression of entropy density and entropy current on the dynamical black hole solution in the higher derivative theories of gravity.  However, this construction works (i.e, it leads to entropy production) only when the amplitude of the dynamics is small, and all terms quadratic or higher order in the amplitude are neglected. Recently it has been extended to quadratic order in amplitude \cite{harveyrecent}.
But clearly, this is not the most satisfying answer; the second law should hold for any dynamics irrespective of its amplitude.
Our final goal is to extend the construction of \cite{our1912,our2105} to the nonlinear orders in amplitude. 

In this note, we have used fluid-gravity correspondence to construct a dual entropy current in the boundary fluid by lifting the entropy current on the horizon via a horizon to boundary map. Since our horizon entropy current works only up to the linear order in the amplitude, we should not expect the fluid entropy current to do any better. So the entropy current constructed in this manner will have non negative divergence only up to the linear order in the dynamical fluid data.

However, in relativistic hydrodynamics we independently know how to extend a given   entropy current that works only up to linear order in amplitude, to an entropy current where the amplitude  is no longer a perturbation parameter \cite{me2012}. So it is reasonable to hope that if we could construct the dual fluid entropy current nonperturbatively and use the  horizon to boundary map in reverse, we might be able to say something about the entropy current in higher derivative theories of gravity in a similar nonperturbative manner.

With this goal in mind, in this note, we have taken the first baby step of constructing the fluid entropy  current dual to the horizon entropy current  \cite{our1912,our2105} in dynamical black holes of Gauss-Bonnet gravity. The fluid entropy current thus constructed depends non trivially on the mapping functions that relate the boundary coordinates with the horizon coordinates. This dependence has complicated our construction since these mapping functions need not admit a derivative expansion like the fluid variables. The immediate future direction would be to search for a particular set of mapping functions so that the final fluid entropy current is expressible only in terms of fluid and fluid-like variables that admit derivative expansion in every stage.\\
 
In this note, we have made a couple of simplifications in this direction. Since both the horizon and the boundary are codimension-one hypersurfaces, naively, there could be $(D-1)$ such mapping functions, where $D$ is the number of bulk dimensions. But using some symmetry and re-arrangement, we could reduce it to only one scalar `non fluid' function, which could be $\phi (x^\mu)$ or $v(x^\mu)$. This scalar is also largely constrained in the sense that if it is specified on a given spatial slice, the consistency equation will fix it everywhere on the horizon (or boundary). So finally, the task of finding appropriate  $(D-1)$ scalar `mapping functions' has been reduced to the search for an appropriate equation, constraining a single scalar on a given spatial slice. \\

In this context, it might be useful to note that the horizon and also the entropy on it have symmetry under the reparametrization of the horizon generator. It has been explored in the case of Einstein-Gauss-Bonnet theory in \cite{Reparametrization_Symmetry, harveyrecent}. The discussion could be extended to include `non-affine' reparametrization of the 
 horizon generators, which might have some direct application for our analysis here.
 

\acknowledgments

We would like to thank Parthajit Biswas, Anirban Dinda and Nilay Kundu for initial collaboration, several useful discussions and many important inputs. S.B. would like to thank Shiraz Minwalla for many valuable discussions. S.R. would like to thank Sourav Dey for helpful discussions. We would also like to acknowledge our debt to the people of India for their steady and generous support to research in the basic sciences.

\appendix
\section{Notation and Identities}\label{app:notiden}
Here, unless explicitly mentioned, all identities and equations are valid only on the horizon, the null hypersurface at $r=0$.
\begin{equation}\label{id:htochi}
\begin{split}
&~h_{ij} = l^\mu_i l^\nu_j \chi_{\mu\nu}\\
&\Gamma_{k,ij} = l^\mu_il^\nu_jl^\alpha_k ~\Gamma_{\alpha,\mu\nu} +\chi_{\mu\nu}~ l^\mu_k(l_i\cdot\partial)l^\nu_j\\
&K_{ij} = l^\mu_i l^\nu_j {\cal K}_{\mu\nu} ~~\text{where}~~{\cal K}_{\mu\nu} = - t^\alpha\Gamma_{\alpha,\mu\nu}
\end{split}
\end{equation}
\underline{Notation related to coordinate transformation}
\begin{equation}\label{not:coord}
\begin{split}
&t^\mu \equiv {\partial x^\mu\over \partial v} \equiv e^\phi n^\mu \equiv e^\phi~ ||n|| ~\hat n^\mu~~~\text{where}~~~n^\mu \equiv G^{\mu r},~~||n|| \equiv \sqrt{n^\mu n^\nu \eta_{\mu\nu}}\\
&\tilde t_\mu = {\partial v\over \partial x^\mu},~~~\tilde t_\mu t^\mu =1,~~~t^\mu \chi_{\mu\nu} =0,~~~t^\mu l^i_{\mu}=\tilde t_{\mu}l^{\mu}_i =0,~~~l^{\mu}_il^j_{\mu}=\delta^j_i,~~~l^{\mu}_il^i_{\nu}+t^{\mu}\tilde{t}_{\nu}=\delta^{\mu}_{\nu}\\
&0= G^{\mu r}G_{rr} + G^{\mu\nu}G_{\nu r}= -\chi^{\mu\nu} u_\nu,~\Rightarrow~~ \chi^{\mu\nu}u_\mu =0\\
&1= G^{rr}G_{rr} + G^{r\mu}G_{\mu r} = -n^\mu u_\mu,~\Rightarrow~~n^\mu u_\mu =-1
\end{split}
\end{equation}\\
\underline{Proof for the first identity in equation \eqref{eq:identities}}\\

Define $\Omega^\mu\equiv \epsilon^{\mu\mu_1\cdots \mu_n} l^{i_1}_{\mu_1}\cdots l^{i_n}_{\mu_n}\left(\epsilon_{i_1\cdots i_n}\over n!\right)$
Now we could show that the expression $ l^{i_1}_{\mu_1}\cdots l^{i_n}_{\mu_n}\epsilon_{i_1\cdots i_n}$ could be expressed as 
$$ l^{i_1}_{\mu_1}\cdots l^{i_n}_{\mu_n}\epsilon_{i_1\cdots i_n}= \Omega^\mu \epsilon_{\mu\mu_1\cdots\mu_n}$$
\begin{equation}
\begin{split}
\Omega^\mu \epsilon_{\mu\mu_1\cdots\mu_n} = \epsilon_{\mu\mu_1\cdots\mu_n} \epsilon^{\mu\nu_1\cdots \nu_n} l^{i_1}_{\nu_1}\cdots l^{i_n}_{\nu_n}\left(\epsilon_{i_1\cdots i_n}\over n!\right)
\end{split}
\end{equation}
\underline{Projectors and related identities}
\begin{equation}\label{id:projectors}
\begin{split}
&{\Delta^\alpha}_\mu \equiv \delta^\alpha_\mu - t^\alpha\tilde t_\mu,~~\text{Note}~~~~\tilde t_\alpha {\Delta^\alpha}_\mu  = {\Delta^\alpha}_\mu t^\mu =0\\
&\bar\chi^{\alpha\beta} = {\Delta^\alpha}_\mu~\chi^{\mu\nu}~{\Delta^\beta}_\nu,~~~~
\bar\chi^{\mu\alpha}\chi_{\alpha\nu} = {\Delta^\mu}_\nu\\
&\chi^{\mu\alpha}\chi_{\alpha\nu} = \delta^\mu_\nu +n^\mu u_\nu~~\Rightarrow~~u_\mu~\chi^{\mu\alpha}\chi_{\alpha\nu} = \chi^{\mu\alpha}\chi_{\alpha\nu} ~n^\nu =0
\end{split}
\end{equation}


\section{First few functions of the coordinate transformation}\label{transformation_functions}
We shall determine $r_{(1)}$ and $x^\mu_{(1)}$ by processing the gauge conditions evaluated at $\rho=0$. On the horizon, the gauge conditions impose the following constraints
\begin{equation}\label{gauge-condition2}
\begin{split}
&-2 u_\mu x^\mu_{(1)} r_{(1)} + x^\mu_{(1)} x^\nu_{(1)} \chi_{\mu\nu} =0,~~~
-u_\mu r_{(1)} t^\mu + t^\nu x^{\mu}_{(1)} \chi_{\mu\nu}=1,~~~
-u_\mu r_{(1)} l^\mu_i + l_i^\nu x^{\mu}_{(1)} \chi_{\mu\nu}=0
\end{split}
\end{equation}
where $$t^\mu \equiv \left(\partial x^\mu_{(0)}\over \partial v\right),~~l^\mu_i \equiv\left(\partial x^\mu_{(0)}\over \partial\alpha_i\right)$$\\
From the second equation using the fact that $t^\mu (\chi_{\mu\nu})_{\rho=0}=0$ we find
\begin{equation}\label{sol1}
\begin{split}
r_{(1)} = -(u_\mu t^\mu)^{-1}
\end{split}
\end{equation}
To simplify the solution for $x^\mu_{(1)}$ we also need the relation between $\chi_{\mu\nu}$ and $h_{ij}$ on the horizon.
\begin{equation}\label{chihij}
\begin{split}
h_{ij}(\rho=0) = &~\left(\partial x^\mu\over\partial \alpha_i\right)\left(\partial x^\nu\over\partial \alpha_j\right)\chi_{\mu\nu}(r=0)= l^\mu_ii^\nu_j\chi_{\mu\nu}\vert_{r=0}\\
h^{ij}(\rho=0) = &~\text{Inverse of $h_{ij}$ at horizon} = \text{$(ij)$ component of the inverse of the bulk metric on the horizon}\\
=&~G^{\rho\mu}\left[ \left(\partial\alpha^i\over\partial x^\mu\right)\left(\partial \alpha^j\over \partial \rho\right)+\left(\partial\alpha^j\over\partial x^\mu\right)\left(\partial \alpha^i\over \partial \rho\right)\right] + G^{\rho\rho}\left(\partial\alpha^j\over\partial \rho\right)\left(\partial \alpha^i\over \partial \rho\right)\\
&+ G^{\mu\nu}\left[\left(\partial\alpha^j\over\partial x^\mu\right)\left(\partial \alpha^i\over \partial x^\nu\right)+\left(\partial\alpha^j\over\partial x^\mu\right)\left(\partial \alpha^i\over \partial x^\nu\right)\right]\\
=&~t^\mu\left[ \left(\partial\alpha^i\over\partial x^\mu\right)\left(\partial \alpha^j\over \partial \rho\right)+\left(\partial\alpha^j\over\partial x^\mu\right)\left(\partial \alpha^i\over \partial \rho\right)\right] + G^{\mu\nu}\left[\left(\partial\alpha^j\over\partial x^\mu\right)\left(\partial \alpha^i\over \partial x^\nu\right)+\left(\partial\alpha^j\over\partial x^\mu\right)\left(\partial \alpha^i\over \partial x^\nu\right)\right]\\
=&~\left(\partial x^\mu\over\partial v\right)\left[ \left(\partial\alpha^i\over\partial x^\mu\right)\left(\partial \alpha^j\over \partial \rho\right)+\left(\partial\alpha^j\over\partial x^\mu\right)\left(\partial \alpha^i\over \partial \rho\right)\right] + G^{\mu\nu}\left[\left(\partial\alpha^j\over\partial x^\mu\right)\left(\partial \alpha^i\over \partial x^\nu\right)+\left(\partial\alpha^j\over\partial x^\mu\right)\left(\partial \alpha^i\over \partial x^\nu\right)\right]\\
=&~ \chi^{\mu\nu}\left[\left(\partial\alpha^j\over\partial x^\mu\right)\left(\partial \alpha^i\over \partial x^\nu\right)+\left(\partial\alpha^j\over\partial x^\mu\right)\left(\partial \alpha^i\over \partial x^\nu\right)\right]= l^i_\mu l^j_\nu \chi^{\mu\nu}
\end{split}
\end{equation}
In the third and the fourth lines, we have used the fact that 
$$G^{\rho\mu}(\rho=0) \propto t^\mu = \left(\partial x^\mu\over\partial v\right)= \text{generator of the horizon}$$
also $G^{\rho\rho}(\rho=0) =0$ and $\chi^{\mu\nu}\equiv G^{\mu\nu} \neq\text{Inverse of $\chi_{\mu\nu}$ (not defined on the horizon)}$.\\
We also need the inverse of these relations i.e., $\chi_{\mu\nu}$ and $\chi^{\mu\nu}$ in terms of $h_{ij}$ etc.
\begin{equation}\label{inverel}
\begin{split}
\chi_{\mu\nu}(r=0) =~& l^i_\mu l^j_\nu h_{ij}\\
\chi^{\mu\nu}(r=0) =~&l^\mu_il^\nu_j h^{ij} + \left[\left(\partial x^\mu\over \partial\rho\right)\left(\partial x^\nu\over \partial\lambda\right) +\left(\partial x^\nu\over \partial\rho\right)\left(\partial x^\mu\over \partial\lambda\right)\right]h^{\rho\lambda}\\
=~&l^\mu_il^\nu_j h^{ij}  + \left[x_{(1)}^\mu t^\nu + x_{(1)}^\nu t^\mu\right] \\
\end{split}
\end{equation}

Now we shall solve for $x^\mu_{(1)}$. 
For convenience, we shall express $x^\mu_{(1)}$ as
\begin{equation}\label{xpara}
x^\mu_{(1)} = P ~t^\mu + P^i~ l^\mu_i
\end{equation}
Substituting \eqref{xpara} and \eqref{sol1} in the third equation of \eqref{gauge-condition2} we find

\begin{equation}\label{xpsil}
\begin{split}
&{u\cdot l_i\over u\cdot t} +P^j~l^\mu_i l^\nu_j\chi_{\mu\nu} =0~\Rightarrow~P^i =-h^{ij} \left(u\cdot l_i\over u\cdot t\right)\\
\text{where}~~~~&h_{ij}(\rho=0) = l^\mu_i l^\nu_j\chi_{\mu\nu},~~~h^{ij} = \text{Inverse of $h_{ij}$}\\
\end{split}
\end{equation}
%
%
%

Now we shall find $P$ from the first equation of \eqref{gauge-condition2}.
\begin{equation}\label{simp5}
\begin{split}
&-2 u_\mu x^\mu_{(1)} r_{(1)} + x^\mu_{(1)} x^\nu_{(1)} \chi_{\mu\nu} =0\\
\Rightarrow~&2P + 2P^i\left(l_i\cdot u\over u\cdot t\right) +x^\mu_{(1)} x^\nu_{(1)} \chi_{\mu\nu}=0\\
\end{split}
\end{equation}
Solving this equation we find $x_{(1)}^\mu$.
\begin{equation}\label{x_mu_1}
\begin{split}
x_{(1)}^\mu = {1\over 2} h^{ij}\left[(u\cdot l_i)(u\cdot l_j)\over (u\cdot t)^2\right] t^\mu - h^{ij} \left[(u\cdot l_i) l^\mu_j\over(u\cdot t)\right]\\
\end{split}
\end{equation}
\vspace{1cm}
{\underline{Some Potentially useful identities for future works}}
\begin{enumerate}
\item $x_{(1)}^\mu$ related
\begin{equation}\label{useful1}
\begin{split}
&x_{(1)}^\mu = {1\over 2} h^{ij}\left[(u\cdot l_i)(u\cdot l_j)\over (u\cdot t)^2\right] t^\mu - h^{ij} \left[(u\cdot l_i) l^\mu_j\over(u\cdot t)\right]\\
&x^\nu_{(1)}\chi_{\mu\nu} = \tilde t_\mu - {u_\mu\over (u\cdot t)}\\
\end{split}
\end{equation}
Using the two identities 
$$h^{ij} = l^i_\mu l^j_\nu \chi^{\mu\nu},~~~l^i_\mu l_i^\nu = \delta^\mu_\nu - t^\mu \tilde t_\nu,~~~\chi^{\mu\nu} u_\nu =0$$
we could further process the expression of $x_{(1)}^\mu$
\begin{equation}\label{process1}
\begin{split}
&h^{ij}(u\cdot l_i)(u\cdot l_j) = (u\cdot t)^2 \left(\tilde t_\alpha\chi^{\alpha\beta}\tilde t_\beta\right)\\
&h^{ij}(l_i\cdot u)l^\mu_j = (u\cdot t)\left[-\tilde t_\nu\chi^{\mu\nu} + t^\mu \left(\tilde t_\alpha\chi^{\alpha\beta}\tilde t_\beta\right)\right]\\
\Rightarrow~&x_{(1)}^\mu =-{1\over 2} \left(\tilde t_\alpha\chi^{\alpha\beta}\tilde t_\beta\right)t^\mu + \tilde t_\nu\chi^{\mu\nu}
\end{split}
\end{equation}

\item Metric related:
\begin{equation}\label{useful2}
\begin{split}
&h_{ij}(\rho=0) = l^\mu_i l^\nu_j \chi_{\mu\nu}(r=0)\\
&h^{ij}(\rho=0) =l^i_\mu l^j_\nu \chi^{\mu\nu}(r=0)\\
&\chi^{\alpha\beta} = l^\alpha_i l^\beta_j h^{ij} + x_{(1)}^\alpha t^\beta + x_{(1)}^\beta t^\alpha\\
\end{split}
\end{equation}


\item Geodesic related
\begin{equation}\label{geo1}
\begin{split}
&t^A\nabla_A t_B \vert_{\rho=0} = 0~~
\Rightarrow~~t^\alpha t^\mu\Gamma_{\alpha,\mu\nu}=0
\end{split}
\end{equation}

\item Extrinsic curvatures
\begin{equation}\label{extrinsic}
\begin{split}
&K_{ij} = l^\mu_i l^\nu_j {\cal K}_{\mu\nu},~~~\bar K_{ij} = l^\mu_i l^\nu_j \bar {\cal K}_{\mu\nu}\\
&\text{where}~~\\
&{\cal K}_{\mu\nu} = -t^\alpha \Gamma_{\alpha,\mu\nu}\\
&\bar {\cal K}_{\mu\nu} = \left(\partial_\mu \tilde t_\nu +\partial_\nu\tilde t_\mu\right) - \left[\partial_\mu u_\nu +\partial_\nu u_\mu\over (u\cdot t)\right] - {\partial_r\chi_{\mu\nu}\over (u\cdot t)} - x_{(1)}^\alpha\Gamma_{\alpha,\mu\nu}\\
&K_{ij}\bar K^{ij} = -\bigg[ \chi^{\mu_1\mu_2}\chi^{\nu_1\nu_2} - \left(\chi^{\mu_1\mu_2}x_{(1)}^{\nu_1}t^{\nu_2}+\chi^{\nu_1\nu_2}x_{(1)}^{\mu_1}t^{\mu_2}\right)\bigg]{\cal K}_{\mu_1\nu_1}\bar {\cal K}_{\mu_2\nu_2}
\end{split}
\end{equation}
%
%
%
\end{enumerate}

\section{Boundary current in terms of fluid variables and $\partial_\mu\phi$}\label{app:decomp}
\subsubsection*{Simplifying $ J^\mu_{space}$}
We shall first show an identity ~~$t^\mu{\cal K}_{\mu\nu}=0$
\begin{equation}\label{id:simpextrin}
\begin{split}
t^\mu{\cal K}_{\mu\nu} = &-t^\mu t^\alpha\Gamma_{\alpha,\mu\nu}\\
=& -t^\mu t^\alpha\left[\partial_\mu\chi_{\nu\alpha}+\partial_\nu\chi_{\mu\alpha} - \partial_\alpha\chi_{\mu\nu}\right]\\
= &-t^\mu t^\alpha\partial_\nu\chi_{\mu\alpha} = -t^\mu\partial_\nu\left[t^\alpha\chi_{\mu\alpha}\right]+t^\mu\chi_{\mu\alpha} \left(\partial_\nu t^\alpha\right) =0
\end{split}
\end{equation}

Now expanding ${\cal D}_\alpha{\cal K}_{\mu\nu}$ we find
\begin{equation}\label{eq:simpextrin2}
\begin{split}
{\cal D}_\alpha{\cal K}_{\mu\nu}=~&\partial_\alpha{\cal K}_{\mu\nu} -\bar\chi^{\theta\phi}\left(\Gamma_{\phi,\alpha\mu}{\cal K}_{\theta\nu}+\Gamma_{\phi,\alpha\nu}{\cal K}_{\theta\mu}\right)+t^\theta \left({\cal K}_{\theta\nu}\partial_\mu\tilde t_\alpha+{\cal K}_{\theta\mu}\partial_\nu\tilde t_\alpha\right)\\
\end{split}
\end{equation}
The last term in the RHS of equation \eqref{eq:simpextrin2} will vanish as a consequence of the identity \eqref{id:simpextrin}. The second term in the RHS of \eqref{eq:simpextrin2} could be further simplified using the expansion of $\bar\chi^{\theta\phi}$.

\begin{equation}\label{eq:approx1}
\begin{split}
&\bar\chi^{\theta\phi}\left(\Gamma_{\phi,\alpha\mu}{\cal K}_{\theta\nu}+\Gamma_{\phi,\alpha\nu}{\cal K}_{\theta\mu}\right)\\
=~&\chi^{\theta\phi}\left(\Gamma_{\phi,\alpha\mu}{\cal K}_{\theta\nu}+\Gamma_{\phi,\alpha\nu}{\cal K}_{\theta\mu}\right)\\
&~-b^{\phi}t^{\theta}\left(\Gamma_{\phi,\alpha\mu}\mathcal{K}_{\theta\nu}+\Gamma_{\phi,\alpha\nu}\mathcal{K}_{\mu\theta}\right)\\
&~+b^{\theta}\left(\mathcal{K}_{\alpha\mu}\mathcal{K}_{\theta\nu}+\mathcal{K}_{\alpha\nu}\mathcal{K}_{\mu\theta}\right)-Bt^{\theta}\left(\mathcal{K}_{\alpha\mu}\mathcal{K}_{\theta\nu}+\mathcal{K}_{\alpha\nu}\mathcal{K}_{\mu\theta}\right)\\
\text{where}~~~~~& b^\mu \equiv \chi^{\mu\nu}\tilde t_\nu,~~~~B\equiv \tilde t_\mu \tilde t_\nu \chi^{\mu\nu}
\end{split}
\end{equation}

Here the term $b^\theta{\cal K}_{\theta\nu}{\cal K}_{\alpha\mu}$ is quadratic in the amplitude of the dynamics and therefore is negligible within our approximation. The last two terms vanish if we apply the identity \eqref{id:simpextrin}. Hence it follows
$${\cal D}_\alpha{\cal K}_{\mu\nu}=~\partial_\alpha{\cal K}_{\mu\nu} -\chi^{\theta\phi}\left(\Gamma_{\phi,\alpha\mu}{\cal K}_{\theta\nu}+\Gamma_{\phi,\alpha\nu}{\cal K}_{\theta\mu}\right) + {\cal O}\left(\epsilon^2\right)$$
From $\partial_{\alpha}{\cal K}_{\mu\nu}$ we can separate the fluid and non-fluid terms in the following way
\begin{equation}
\begin{split}
\partial_{\alpha}{\cal K}_{\mu\nu}=-e^{\phi}\left\{\left(\partial_{\alpha}n^{\nu_1}\right)\Gamma_{\nu_1,\mu\nu}+n^{\nu_1}\partial_{\alpha}\Gamma_{\nu_1,\mu\nu}\right\}-e^{\phi}\left(\partial_{\alpha}\phi\right)n^{\nu_1}\Gamma_{\nu_1,\mu\nu}
\end{split}
\end{equation}
Now for convenience we will write the expression for  $ J^\mu_{space}$ as a sum of two terms
\begin{equation}
\begin{split}
J^\mu_{space}=T_1+T_2
\end{split}
\end{equation}
with
\begin{equation}
\begin{split}
&T_1=-4\alpha^2\frac{1}{\sqrt{g^{(b)}}}\frac{\sqrt{H}}{\sqrt{t^\alpha t^\beta g^{(b)}_{\alpha\beta}}}\left(\bar\chi^{\gamma\alpha}\bar\chi^{\mu\beta} - \bar\chi^{\gamma\mu}\bar\chi^{\alpha\beta}\right)\left(\partial_{\gamma}{\cal K}_{\alpha\beta}\right)\\
&T_2=4\alpha^2\frac{1}{\sqrt{g^{(b)}}}\frac{\sqrt{H}}{\sqrt{t^\alpha t^\beta g^{(b)}_{\alpha\beta}}}\left(\bar\chi^{\gamma\alpha}\bar\chi^{\mu\beta} - \bar\chi^{\gamma\mu}\bar\chi^{\alpha\beta}\right)\chi^{\theta\phi}\left(\Gamma_{\phi,\gamma\alpha}{\cal K}_{\theta\beta}+\Gamma_{\phi,\gamma\beta}{\cal K}_{\theta\alpha}\right)
\end{split}
\end{equation}
Now we use the identity of \eqref{id:simpextrin} to simplify the terms and \eqref{eq:tildet} to separate the terms
\begin{equation}
\begin{split}
\left[T_1\right]_{fluid}=&4\alpha^2\frac{1}{\sqrt{g^{(b)}}}\frac{\sqrt{H}}{\sqrt{n^\alpha n^\beta g^{(b)}_{\alpha\beta}}}\left\{\left(\partial_\gamma n^{\nu_1}\right)\Gamma_{\nu_1,\alpha\beta}+n^{\nu_1}\partial_\gamma \Gamma_{\nu_1,\alpha\beta}\right\}\bigg[\left(\chi^{\alpha\alpha_1}\partial_{\alpha_1}L\right)\left(\chi^{\beta\beta_1}\partial_{\beta_1}L\right)n^\gamma n^\mu\\
&-\left(\chi^{\beta\beta_1}\partial_{\beta_1}L\right)n^\mu \chi^{\gamma\alpha}-\left(\chi^{\alpha\alpha_1}\partial_{\alpha_1}L\right)\left(\chi^{\gamma\gamma_1}\partial_{\gamma_1}L\right)n^\beta n^\mu+\left(\chi^{\gamma\gamma_1}\partial_{\gamma_1}L\right)n^\mu \chi^{\alpha\beta}+\left(\chi^{\alpha\alpha_1}\partial_{\alpha_1}L\right)n^\beta \chi^{\gamma\mu}\\
&-\chi^{\alpha\beta}\chi^{\gamma\mu}+\left(\chi^{\beta\beta_1}\partial_{\beta_1}L\right)n^\alpha \chi^{\gamma\mu}-\left(\chi^{\alpha\alpha_1}\partial_{\alpha_1}L\right)n^\gamma \chi^{\mu\beta}+\chi^{\gamma\alpha}\chi^{\mu\beta}-\left(\chi^{\gamma\gamma_1}\partial_{\gamma_1}L\right)n^\alpha \chi^{\mu\beta}\\
&+\left(\chi^{\mu\mu_1}\partial_{\mu_1}L\right)n^\gamma \chi^{\alpha\beta}-\left(\chi^{\beta\beta_1}\partial_{\beta_1}L\right)\left(\chi^{\mu\mu_1}\partial_{\mu_1}L\right)n^\alpha n^\gamma-\left(\chi^{\mu\mu_1}\partial_{\mu_1}L\right)n^\beta \chi^{\gamma\alpha}\\
+&\left(\chi^{\gamma\gamma_1}\partial_{\gamma_1}L\right)\left(\chi^{\mu\mu_1}\partial_{\mu_1}L\right)n^\alpha n^\beta+\left(\chi^{\sigma_1\sigma_2}\partial_{\sigma_1}L~\partial_{\sigma_2}L\right)\big(n^\beta n^\mu \chi^{\gamma\alpha}-n^\gamma n^\mu \chi^{\alpha\beta}-n^\alpha n^\beta \chi^{\gamma\mu}+n^\alpha n^\gamma \chi^{\mu\beta}\big)\bigg]
\end{split}
\end{equation}
\begin{equation}
\begin{split}
\left[T_1\right]_{non-fluid}&=4\alpha^2\frac{1}{\sqrt{g^{(b)}}}\frac{\sqrt{H}}{\sqrt{n^\alpha n^\beta g^{(b)}_{\alpha\beta}}}\left\{\left(\partial_\gamma \phi\right)n^{\nu_1}\Gamma_{\nu_1,\alpha\beta}\right\}\bigg[\left(\chi^{\alpha\alpha_1}\partial_{\alpha_1}L\right)\left(\chi^{\beta\beta_1}\partial_{\beta_1}L\right)n^\gamma n^\mu\\
&-\left(\chi^{\beta\beta_1}\partial_{\beta_1}L\right)n^\mu \chi^{\gamma\alpha}-\left(\chi^{\alpha\alpha_1}\partial_{\alpha_1}L\right)\left(\chi^{\gamma\gamma_1}\partial_{\gamma_1}L\right)n^\beta n^\mu+\left(\chi^{\gamma\gamma_1}\partial_{\gamma_1}L\right)n^\mu \chi^{\alpha\beta}+\left(\chi^{\alpha\alpha_1}\partial_{\alpha_1}L\right)n^\beta \chi^{\gamma\mu}\\
&-\chi^{\alpha\beta}\chi^{\gamma\mu}+\left(\chi^{\beta\beta_1}\partial_{\beta_1}L\right)n^\alpha \chi^{\gamma\mu}-\left(\chi^{\alpha\alpha_1}\partial_{\alpha_1}L\right)n^\gamma \chi^{\mu\beta}+\chi^{\gamma\alpha}\chi^{\mu\beta}-\left(\chi^{\gamma\gamma_1}\partial_{\gamma_1}L\right)n^\alpha \chi^{\mu\beta}\\
&+\left(\chi^{\mu\mu_1}\partial_{\mu_1}L\right)n^\gamma \chi^{\alpha\beta}-\left(\chi^{\beta\beta_1}\partial_{\beta_1}L\right)\left(\chi^{\mu\mu_1}\partial_{\mu_1}L\right)n^\alpha n^\gamma-\left(\chi^{\mu\mu_1}\partial_{\mu_1}L\right)n^\beta \chi^{\gamma\alpha}\\
+&\left(\chi^{\gamma\gamma_1}\partial_{\gamma_1}L\right)\left(\chi^{\mu\mu_1}\partial_{\mu_1}L\right)n^\alpha n^\beta+\left(\chi^{\sigma_1\sigma_2}\partial_{\sigma_1}L~\partial_{\sigma_2}L\right)\big(n^\beta n^\mu \chi^{\gamma\alpha}-n^\gamma n^\mu \chi^{\alpha\beta}-n^\alpha n^\beta \chi^{\gamma\mu}+n^\alpha n^\gamma \chi^{\mu\beta}\big)\bigg]\\
&+4\alpha^2\frac{1}{\sqrt{g^{(b)}}}\frac{\sqrt{H}}{\sqrt{n^\alpha n^\beta g^{(b)}_{\alpha\beta}}}\left\{\left(\partial_\gamma \phi\right)n^{\nu_1}\Gamma_{\nu_1,\alpha\beta}+\left(\partial_\gamma n^{\nu_1}\right)\Gamma_{\nu_1,\alpha\beta}+n^{\nu_1}\partial_\gamma \Gamma_{\nu_1,\alpha\beta}\right\}\bigg[\big(-L~\partial_{\alpha_1}L~\partial_{\beta_1}\phi\\
&-L~\partial_{\beta_1}L~\partial_{\alpha_1}\phi+L^2~\partial_{\alpha_1}\phi~\partial_{\beta_1}\phi\big)n^\gamma n^\mu \chi^{\alpha\alpha_1}\chi^{\beta\beta_1}+L\left(\chi^{\beta\beta_1}\partial_{\beta_1}\phi\right)n^\mu \chi^{\gamma\alpha}-\big(-L~\partial_{\alpha_1}L~\partial_{\gamma_1}\phi\\
&-L~\partial_{\gamma_1}L~\partial_{\alpha_1}\phi+L^2~\partial_{\alpha_1}\phi~\partial_{\gamma_1}\phi\big)n^\beta n^\mu \chi^{\alpha\alpha_1}\chi^{\gamma\gamma_1}-L\left(\chi^{\gamma\gamma_1}\partial_{\gamma_1}\phi\right)n^\mu \chi^{\alpha\beta}-L\left(\chi^{\alpha\alpha_1}\partial_{\alpha_1}\phi\right)n^\beta \chi^{\gamma\mu}\\
&-L\left(\chi^{\beta\beta_1}\partial_{\beta_1}\phi\right)n^\alpha \chi^{\gamma\mu}+L\left(\chi^{\alpha\alpha_1}\partial_{\alpha_1}\phi\right)n^\gamma \chi^{\mu\beta}+L\left(\chi^{\gamma\gamma_1}\partial_{\gamma_1}\phi\right)n^\alpha \chi^{\mu\beta}-L\left(\chi^{\mu\mu_1}\partial_{\mu_1}\phi\right)n^\gamma \chi^{\alpha\beta}\\
&-\big(-L~\partial_{\beta_1}L~\partial_{\mu_1}\phi-L~\partial_{\mu_1}L~\partial_{\beta_1}\phi+L^2~\partial_{\beta_1}\phi~\partial_{\mu_1}\phi\big)n^\alpha n^\gamma \chi^{\beta\beta_1}\chi^{\mu\mu_1}+L\left(\chi^{\mu\mu_1}\partial_{\mu_1}\phi\right)n^\beta \chi^{\gamma\alpha}\\
&+\big(-L~\partial_{\gamma_1}L~\partial_{\mu_1}\phi-L~\partial_{\gamma_1}\phi~\partial_{\mu_1}L+L^2~\partial_{\gamma_1}\phi~\partial_{\mu_1}\phi\big)n^\alpha n^\beta \chi^{\gamma\gamma_1}\chi^{\mu\mu_1}-\big(L~\partial_{\sigma_1}L~\partial_{\sigma_2}\phi\\
&+L~\partial_{\sigma_1}\phi~\partial_{\sigma_2}L-L^2~\partial_{\sigma_1}\phi~\partial_{\sigma_2}\phi\big)\chi^{\sigma_1\sigma_2}\big(n^\beta n^\mu \chi^{\gamma\alpha}-n^\gamma n^\mu \chi^{\alpha\beta}-n^\alpha n^\beta \chi^{\gamma\mu}+n^\alpha n^\gamma \chi^{\mu\beta}\big)\bigg]
\end{split}
\end{equation}

\begin{equation}
\begin{split}
\left[T_2\right]_{fluid}=&4\alpha^2\frac{1}{\sqrt{g^{(b)}}}\frac{\sqrt{H}}{\sqrt{n^\alpha n^\beta g^{(b)}_{\alpha\beta}}}\chi^{\theta\phi}\bigg[\left(n^{\sigma_3}\Gamma_{\sigma_3,\theta\beta}\Gamma_{\phi,\gamma\alpha}+n^{\sigma_4}\Gamma_{\sigma_4,\theta\alpha}\Gamma_{\phi,\gamma\beta}\right)\bigg\{-\left(\chi^{\alpha\alpha_1}\partial_{\alpha_1}L\right)\left(\chi^{\beta\beta_1}\partial_{\beta_1}L\right)n^{\mu}n^{\gamma}\\
&+\left(\chi^{\beta\beta_1}\partial_{\beta_1}L\right)n^{\mu}\chi^{\gamma\alpha}-\left(\chi^{\gamma\gamma_1}\partial_{\gamma_1}L\right)n^{\mu}\chi^{\alpha\beta}+\chi^{\alpha\beta}\chi^{\gamma\mu}+\left(\chi^{\alpha\alpha_1}\partial_{\alpha_1}L\right)n^{\gamma}\chi^{\mu\beta}-\chi^{\gamma\alpha}\chi^{\mu\beta}\\
&-\left(\chi^{\mu\mu_1}\partial_{\mu_1}L\right)n^{\gamma}\chi^{\alpha\beta}+\left(\chi^{\sigma_1\sigma_2}\partial_{\sigma_1}L~\partial_{\sigma_2}L\right)n^{\gamma}n^{\mu}\chi^{\alpha\beta}\bigg\}\\
&+\left(n^{\sigma_4}\Gamma_{\sigma_4,\theta\alpha}\right)\Gamma_{\phi,\gamma\beta}\bigg\{\left(\chi^{\alpha\alpha_1}\partial_{\alpha_1}L\right)\left(\chi^{\gamma\gamma_1}\partial_{\gamma_1}L\right)n^{\beta}n^{\mu}-\left(\chi^{\alpha\alpha_1}\partial_{\alpha_1}L\right)n^{\beta}\chi^{\gamma\mu}+\left(\chi^{\mu\mu_1}\partial_{\mu_1}L\right)n^{\beta}\chi^{\gamma\alpha}\\
&-\left(\chi^{\sigma_1\sigma_2}\partial_{\sigma_1}L~\partial_{\sigma_2}L\right)n^{\beta}n^{\mu}\chi^{\gamma\alpha}\bigg\}+\left(n^{\sigma_3}\Gamma_{\sigma_3,\theta\beta}\right)\Gamma_{\phi,\gamma\alpha}\bigg\{-\left(\chi^{\beta\beta_1}\partial_{\beta_1}L\right)n^{\alpha}\chi^{\gamma\mu}+\left(\chi^{\gamma\gamma_1}\partial_{\gamma_1}L\right)n^{\alpha}\chi^{\mu\beta}\\
&+\left(\chi^{\mu\mu_1}\partial_{\mu_1}L\right)\left(\chi^{\beta\beta_1}\partial_{\beta_1}L\right)n^\alpha n^\gamma-\left(\chi^{\sigma_1\sigma_2}\partial_{\sigma_1}L~\partial_{\sigma_2}L\right)n^{\alpha}n^{\gamma}\chi^{\mu\beta}\bigg\}\bigg]
\end{split}
\end{equation}
\begin{equation}
\begin{split}
\left[T_2\right]_{non-fluid}=&4\alpha^2\frac{1}{\sqrt{g^{(b)}}}\frac{\sqrt{H}}{\sqrt{n^\alpha n^\beta g^{(b)}_{\alpha\beta}}}\chi^{\theta\phi}\bigg[\bigg\{\left(L~\partial_{\alpha_1}L~\partial_{\beta_1}\phi+L~\partial_{\alpha_1}\phi~\partial_{\beta_1}L-L^2~\partial_{\alpha_1}\phi~\partial_{\beta_1}\phi\right)n^{\mu}n^{\gamma}\chi^{\alpha\alpha_1}\chi^{\beta\beta_1}\\
&+L\left(\chi^{\gamma\gamma_1}\partial_{\gamma_1}\phi~\chi^{\alpha\beta}-\chi^{\beta\beta_1}\partial_{\beta_1}\phi~\chi^{\gamma\alpha}\right)n^{\mu}-L\left(\chi^{\alpha\alpha_1}\partial_{\alpha_1}\phi~\chi^{\mu\beta}\right)n^{\gamma}\\
&+L\left(\chi^{\mu\mu_1}\partial_{\mu_1}\phi~\chi^{\alpha\beta}\right)n^{\gamma}\bigg\}\left(n^{\sigma_3}\Gamma_{\sigma_3,\theta\beta}\Gamma_{\phi,\gamma\alpha}+n^{\sigma_4}\Gamma_{\sigma_4,\theta\alpha}\Gamma_{\phi,\gamma\beta}\right)\\
&-n^\beta n^\mu\left(n^{\sigma_4}\Gamma_{\sigma_4,\theta\alpha}\Gamma_{\phi,\gamma\beta}\right)\chi^{\alpha\alpha_1}\chi^{\beta\beta_1} \left(L~\partial_{\alpha_1}L~\partial_{\gamma_1}\phi+L~\partial_{\alpha_1}\phi~\partial_{\gamma_1}L-L^2~\partial_{\alpha_1}\phi~\partial_{\gamma_1}\phi\right)\\
&+L\left(n^{\sigma_4}\Gamma_{\sigma_4,\theta\alpha}\right)\Gamma_{\phi,\gamma\beta}\left\{\left(\chi^{\alpha\alpha_1}\partial_{\alpha_1}\phi\right)~n^{\beta}\chi^{\gamma\mu}-\left(\chi^{\mu\mu_1}\partial_{\mu_1}\phi\right)n^\beta \chi^{\gamma\alpha}\right\}\\
&+L\left(n^{\sigma_3}\Gamma_{\sigma_3,\theta\beta}\right)\Gamma_{\phi,\gamma\alpha}\left\{\chi^{\beta\beta_1}\partial_{\beta_1}\phi~n^{\alpha}\chi^{\gamma\mu}-\chi^{\gamma\gamma_1}\partial_{\gamma_1}\phi~n^{\alpha}\chi^{\mu\beta}\right\}\\
&-\left(L~\partial_{\beta_1}L~\partial_{\mu_1}\phi+L~\partial_{\beta_1}\phi~\partial_{\mu_1}L-L^2~\partial_{\beta_1}\phi~\partial_{\mu_1}\phi\right)n^{\alpha}n^{\gamma}\chi^{\mu\mu_1}\chi^{\beta\beta_1}\left(n^{\sigma_3}\Gamma_{\sigma_3,\theta\beta}\right)\Gamma_{\phi,\gamma\alpha}\\
&-\chi^{\sigma_1\sigma_2}\left(L~\partial_{\sigma_1}L~\partial_{\sigma_2}\phi+L~\partial_{\sigma_1}\phi~\partial_{\sigma_2}L-L^2~\partial_{\sigma_1}\phi~\partial_{\sigma_2}\phi\right)\bigg\{-\left(n^{\sigma_4}\Gamma_{\sigma_4,\theta\alpha}\right)\Gamma_{\phi,\gamma\beta}n^{\beta}n^{\mu}\chi^{\gamma\alpha}\\
&-\left(n^{\sigma_3}\Gamma_{\sigma_3,\theta\beta}\right)\Gamma_{\phi,\gamma\alpha}n^{\alpha}n^{\gamma}\chi^{\mu\beta}+\left(n^{\sigma_3}\Gamma_{\sigma_3,\theta\beta}\Gamma_{\phi,\gamma\alpha}+n^{\sigma_4}\Gamma_{\sigma_4,\theta\alpha}\Gamma_{\phi,\gamma\beta}\right)n^{\gamma}n^{\mu}\chi^{\alpha\beta}\bigg\}\bigg]
\end{split}
\end{equation}

\subsubsection*{Simplifying $J^\mu_{time}$}
In this section we will write down the intrinsic Ricci scalar as a sum of `fluid' and `non-fluid' terms.\\\
Using the definition of ${\cal K}$ and ignoring the terms quadratic in the amplitude of dynamics, we can write
\begin{equation}
\begin{split}
{\cal R} = \left(\bar\chi^{\mu_1\nu_1}\bar\chi^{\mu_2\nu_2}-\bar\chi^{\mu_1\nu_2}\bar\chi^{\mu_2\nu_1}\right)\bigg[&\partial_{\mu_1}\Gamma_{\nu_1,\mu_2\nu_2} - \chi^{\alpha_1\alpha_2}\Gamma_{\alpha_1,\mu_1\nu_1}\Gamma_{\alpha_2,\mu_2\nu_2} - b^{\alpha_1}\Gamma_{\alpha_1,\mu_1\nu_1}{\cal K}_{\mu_2\nu_2}- b^{\alpha_2}\Gamma_{\alpha_2,\mu_2\nu_2}{\cal K}_{\mu_1\nu_1}\\
&+2 {\cal K}_{\mu_1\nu_1}\left(\partial_{\mu_2}\tilde t_{\nu_2}\right)\bigg]\\
=T_1+T_2+T_3+T_4+T_5
\end{split}
\end{equation}
Now we use the identity of \eqref{id:simpextrin} to simplify the terms and \eqref{eq:tildet} to separate the terms
\begin{equation}
\begin{split}
[T_1+T_2]_{fluid}=&\bigg[\partial_{\mu_1}\Gamma_{\nu_1,\mu_2\nu_2} - \chi^{\alpha_1\alpha_2}\Gamma_{\alpha_1,\mu_1\nu_1}\Gamma_{\alpha_2,\mu_2\nu_2}\bigg]\bigg[-\left(\chi^{\mu_2\theta_2}\partial_{\theta_2}L\right)n^{\nu_2}\chi^{\mu_1\nu_1}+\left(\chi^{\mu_2\theta_2}\partial_{\theta_2}L\right)n^{\nu_1}\chi^{\mu_1\nu_2}\\
&+\left(\chi^{\mu_1\theta_3}\partial_{\theta_3}L\right)n^{\nu_2}\chi^{\mu_2\nu_1}-\left(\chi^{\mu_1\theta_3}\partial_{\theta_3}L\right)n^{\nu_1}\chi^{\mu_2\nu_2}-\chi^{\mu_1\nu_2}\chi^{\mu_2\nu_1}+\chi^{\mu_1\nu_1}\chi^{\mu_2\nu_2}\\
&+\left(\chi^{\alpha\beta}\partial_{\alpha}L~\partial_{\beta}L\right)\left(n^{\mu_2}n^{\nu_2}\chi^{\mu_1\nu_1}-n^{\mu_2}n^{\nu_1}\chi^{\mu_1\nu_2}-n^{\mu_1}n^{\nu_2}\chi^{\mu_2\nu_1}+n^{\mu_1}n^{\nu_1}\chi^{\mu_2\nu_2}\right)\\
&-\left(\chi^{\mu_1\theta_3}\partial_{\theta_3}L\right)\left(\chi^{\nu_1\theta_4}\partial_{\theta_4}L\right)n^{\mu_2}n^{\nu_2}+\left(\chi^{\nu_1\theta_4}\partial_{\theta_4}L\right)\left(\chi^{\mu_2\theta_2}\partial_{\theta_2}L\right)n^{\mu_1}n^{\nu_2}\\
&+\left(\chi^{\nu_1\theta_4}\partial_{\theta_4}L\right)n^{\mu_2}\chi^{\mu_1\nu_2}-\left(\chi^{\nu_1\theta_4}\partial_{\theta_4}L\right)n^{\mu_1}\chi^{\mu_2\nu_2}+\left(\chi^{\nu_2\theta_5}\partial_{\theta_5}L\right)\bigg\{\left(\chi^{\mu_1\theta_3}\partial_{\theta_3}L\right)n^{\mu_2}n^{\nu_1}\\
&-\left(\chi^{\mu_2\theta_2}\partial_{\theta_2}L\right)n^{\mu_1}n^{\nu_1}-\chi^{\mu_1\nu_1}n^{\mu_2}+\chi^{\mu_2\nu_1}n^{\mu_1}\bigg\}\bigg]\\
\end{split}
\end{equation}

\begin{equation}
\begin{split}
[T_1+T_2]_{non-fluid}=&\bigg[\partial_{\mu_1}\Gamma_{\nu_1,\mu_2\nu_2} - \chi^{\alpha_1\alpha_2}\Gamma_{\alpha_1,\mu_1\nu_1}\Gamma_{\alpha_2,\mu_2\nu_2}\bigg]\bigg[L\left(\chi^{\mu_2\theta_2}\partial_{\theta_2}\phi\right)\left(n^{\nu_2}\chi^{\mu_1\nu_1}-n^{\nu_1}\chi^{\mu_1\nu_2}\right)\\
&+L\left(\chi^{\mu_1\theta_3}\partial_{\theta_3}\phi\right)\left(n^{\nu_1}\chi^{\mu_2\nu_2}-n^{\nu_2}\chi^{\mu_2\nu_1}\right)+L\left(\chi^{\nu_1\theta_4}\partial_{\theta_4}\phi\right)\left(n^{\mu_1}\chi^{\mu_2\nu_2}-n^{\mu_2}\chi^{\mu_1\nu_2}\right)\\
&+\left\{n^{\mu_2}n^{\nu_2}\chi^{\mu_1\nu_1}-n^{\mu_2}n^{\nu_1}\chi^{\mu_1\nu_2}-n^{\mu_1}n^{\nu_2}\chi^{\mu_2\nu_1}+n^{\mu_1}n^{\nu_1}\chi^{\mu_2\nu_2}\right\}\bigg\{-L\left(\chi^{\alpha\beta}\partial_{\alpha}L~\partial_{\beta}\phi\right)\\
&-L\left(\chi^{\alpha\beta}\partial_{\alpha}\phi~\partial_{\beta}L\right)+L^2\left(\chi^{\alpha\beta}\partial_{\alpha}\phi~\partial_{\beta}\phi\right)\bigg\}+L\left(\chi^{\mu_1\theta_3}\partial_{\theta_3}\phi\right)\left(\chi^{\nu_1\theta_4}\partial_{\theta_4}L\right)n^{\mu_2}n^{\nu_2}\\
&+L\left(\chi^{\nu_1\theta_4}\partial_{\theta_4}\phi\right)\left(\chi^{\mu_1\theta_3}\partial_{\theta_3}L\right)n^{\mu_2}n^{\nu_2}-L^2\left(\chi^{\mu_1\theta_3}\partial_{\theta_3}\phi\right)\left(\chi^{\nu_1\theta_4}\partial_{\theta_4}\phi\right)n^{\mu_2}n^{\nu_2}\\
&-L\left(\chi^{\mu_2\theta_2}\partial_{\theta_2}\phi\right)\left(\chi^{\nu_1\theta_4}\partial_{\theta_4}L\right)n^{\mu_1}n^{\nu_2}-L\left(\chi^{\nu_1\theta_4}\partial_{\theta_4}\phi\right)\left(\chi^{\mu_2\theta_2}\partial_{\theta_2}L\right)n^{\mu_1}n^{\nu_2}\\
&+L^2\left(\chi^{\nu_1\theta_4}\partial_{\theta_4}\phi\right)\left(\chi^{\mu_2\theta_2}\partial_{\theta_2}\phi\right)n^{\mu_1}n^{\nu_2}-L\left(\chi^{\nu_2\theta_5}\partial_{\theta_5}\phi\right)\bigg\{-n^{\mu_2}\chi^{\mu_1\nu_1}+n^{\mu_1}\chi^{\mu_2\nu_1}\\
&+\left(\chi^{\mu_1\theta_3}\partial_{\theta_3}	L\right)n^{\mu_2}n^{\nu_1}-\left(\chi^{\mu_2\theta_2}\partial_{\theta_2}L\right)n^{\mu_1}n^{\nu_1}-L\left(\chi^{\mu_1\theta_3}\partial_{\theta_3}\phi\right)n^{\mu_2}n^{\nu_1}\\
&+L\left(\chi^{\mu_2\theta_2}\partial_{\theta_2}\phi\right)n^{\mu_1}n^{\nu_1}\bigg\}\bigg]\\
\end{split}
\end{equation}

\begin{equation}
\begin{split}
[T_3]_{fluid}=&\left(\chi^{\alpha_1\theta_6}\partial_{\theta_6}L\right)\left(n^{\sigma_3}\Gamma_{\sigma_3,\mu_2\nu_2}\right)\Gamma_{\alpha_1,\mu_1\nu_1}\bigg[\left(\chi^{\mu_2\theta_2}\partial_{\theta_2}L\right)n^{\nu_1} \chi^{\mu_1\nu_2}-\left(\chi^{\mu_1\theta_3}\partial_{\theta_3}L\right)n^{\nu_1} \chi^{\mu_2\nu_2}\\
&-\left(\chi^{\nu_1\theta_4}\partial_{\theta_4}L\right)n^{\mu_1} \chi^{\mu_2\nu_2}+\left(\chi^{\nu_2\theta_5}\partial_{\theta_5}L\right)n^{\mu_1} \chi^{\mu_2\nu_1}+\chi^{\mu_1\nu_1}\chi^{\mu_2\nu_2}-\chi^{\mu_1\nu_2}\chi^{\mu_2\nu_1}\\
&+\left(\chi^{\alpha\beta}\partial_{\alpha}L~\partial_{\beta}L\right)n^{\mu_1}n^{\nu_1}\chi^{\mu_2\nu_2}-\left(\chi^{\mu_2\theta_2}\partial_{\theta_2}L\right)\left(\chi^{\nu_2\theta_5}\partial_{\theta_5}L\right)n^{\mu_1}n^{\nu_1}\bigg]\\
\end{split}
\end{equation}

\begin{equation}
\begin{split}
[T_3]_{non-fluid}=&\left(\chi^{\alpha_1\theta_6}\partial_{\theta_6}L\right)\left(n^{\sigma_3}\Gamma_{\sigma_3,\mu_2\nu_2}\right)\Gamma_{\alpha_1,\mu_1\nu_1}\bigg[-L\left(\chi^{\mu_2\theta_2}\partial_{\theta_2}\phi\right)n^{\nu_1} \chi^{\mu_1\nu_2}+L\left(\chi^{\nu_1\theta_4}\partial_{\theta_4}\phi\right)n^{\mu_1} \chi^{\mu_2\nu_2}\\
&+L\left(\chi^{\mu_1\theta_3}\partial_{\theta_3}\phi\right)n^{\nu_1} \chi^{\mu_2\nu_2}-L\left(\chi^{\nu_2\theta_5}\partial_{\theta_5}\phi\right)n^{\mu_1} \chi^{\mu_2\nu_1}-\big(L~\partial_{\alpha}L~\partial_{\beta}\phi+L~\partial_{\alpha}\phi~\partial_{\beta}L\\
&-L^2~\partial_{\alpha}\phi~\partial_{\beta}\phi\big)\chi^{\alpha\beta}n^{\mu_1}n^{\nu_1}\chi^{\mu_2\nu_2}\bigg]\\
&-L\left(\chi^{\alpha_1\theta_6}\partial_{\theta_6}\phi\right)\left(n^{\sigma_3}\Gamma_{\sigma_3,\mu_2\nu_2}\right)\Gamma_{\alpha_1,\mu_1\nu_1}\bigg[-L\left(\chi^{\mu_2\theta_2}\partial_{\theta_2}\phi\right)n^{\nu_1} \chi^{\mu_1\nu_2}+L\left(\chi^{\nu_1\theta_4}\partial_{\theta_4}\phi\right)n^{\mu_1} \chi^{\mu_2\nu_2}\\
&+L\left(\chi^{\mu_1\theta_3}\partial_{\theta_3}\phi\right)n^{\nu_1} \chi^{\mu_2\nu_2}-L\left(\chi^{\nu_2\theta_5}\partial_{\theta_5}\phi\right)n^{\mu_1} \chi^{\mu_2\nu_1}-\big(L~\partial_{\alpha}L~\partial_{\beta}\phi+L~\partial_{\alpha}\phi~\partial_{\beta}L\\
&-L^2~\partial_{\alpha}\phi~\partial_{\beta}\phi\big)\chi^{\alpha\beta}n^{\mu_1}n^{\nu_1}\chi^{\mu_2\nu_2}+\left(\chi^{\mu_2\theta_2}\partial_{\theta_2}L\right)n^{\nu_1} \chi^{\mu_1\nu_2}-\left(\chi^{\mu_1\theta_3}\partial_{\theta_3}L\right)n^{\nu_1} \chi^{\mu_2\nu_2}\\
&-\left(\chi^{\nu_1\theta_4}\partial_{\theta_4}L\right)n^{\mu_1} \chi^{\mu_2\nu_2}+\left(\chi^{\nu_2\theta_5}\partial_{\theta_5}L\right)n^{\mu_1} \chi^{\mu_2\nu_1}+\chi^{\mu_1\nu_1}\chi^{\mu_2\nu_2}-\chi^{\mu_1\nu_2}\chi^{\mu_2\nu_1}\\
&+\left(\chi^{\alpha\beta}\partial_{\alpha}L~\partial_{\beta}L\right)n^{\mu_1}n^{\nu_1}\chi^{\mu_2\nu_2}-\left(\chi^{\mu_2\theta_2}\partial_{\theta_2}L\right)\left(\chi^{\nu_2\theta_5}\partial_{\theta_5}L\right)n^{\mu_1}n^{\nu_1}\bigg]\\
\end{split}
\end{equation}

\begin{equation}
\begin{split}
\left[T_4+T_5\right]_{fluid}=&n^{\sigma_4}\Gamma_{\sigma_4,\mu_1\nu_1}\left(2\partial_{\mu_2}\partial_{\nu_2}L-\chi^{\alpha_2\theta_7}\partial_{\theta_7}L~\Gamma_{\alpha_2,\mu_2\nu_2}\right)\bigg[-\left(\chi^{\alpha\beta}\partial_{\alpha}L~\partial_{\beta}L\right)n^{\mu_2}n^{\nu_2}\chi^{\mu_1\nu_1}\\
&+\left(\chi^{\mu_2\theta_2}\partial_{\theta_2}L\right)n^{\nu_2} \chi^{\mu_1\nu_1}-\left(\chi^{\mu_1\theta_3}\partial_{\theta_3}L\right)n^{\nu_2} \chi^{\mu_2\nu_1}-\left(\chi^{\nu_1\theta_4}\partial_{\theta_4}L\right)n^{\mu_2} \chi^{\mu_1\nu_2}\\
&+\left(\chi^{\nu_2\theta_5}\partial_{\theta_5}L\right)n^{\mu_2} \chi^{\mu_1\nu_1}+\left(\chi^{\mu_1\theta_3}\partial_{\theta_3}L\right)\left(\chi^{\nu_1\theta_4}\partial_{\theta_4}L\right)n^{\mu_2}n^{\nu_2}+\chi^{\mu_1\nu_2}\chi^{\mu_2\nu_1}-\chi^{\mu_1\nu_1}\chi^{\mu_2\nu_2}\bigg]\\
\end{split}
\end{equation}

\begin{equation}
\begin{split}
\left[T_4+T_5\right]_{non-fluid}=&n^{\sigma_4}\Gamma_{\sigma_4,\mu_1\nu_1}\left(2\partial_{\mu_2}\partial_{\nu_2}L-\chi^{\alpha_2\theta_7}\partial_{\theta_7}L~\Gamma_{\alpha_2,\mu_2\nu_2}\right)\bigg[L\left(\chi^{\alpha\beta}\partial_{\alpha}\phi~\partial_{\beta}L\right)n^{\mu_2}n^{\nu_2}\chi^{\mu_1\nu_1}\\
&-L^2\left(\chi^{\alpha\beta}\partial_{\alpha}\phi~\partial_{\beta}\phi\right)n^{\mu_2}n^{\nu_2}\chi^{\mu_1\nu_1}+L\left(\chi^{\alpha\beta}\partial_{\alpha}L~\partial_{\beta}\phi\right)n^{\mu_2}n^{\nu_2}\chi^{\mu_1\nu_1}\\
&-L\left(\chi^{\mu_2\theta_2}\partial_{\theta_2}\phi\right)n^{\nu_2} \chi^{\mu_1\nu_1}+L\left(\chi^{\mu_1\theta_3}\partial_{\theta_3}\phi\right)n^{\nu_2} \chi^{\mu_2\nu_1}+L\left(\chi^{\nu_1\theta_4}\partial_{\theta_4}\phi\right)n^{\mu_2} \chi^{\mu_1\nu_2}\\
&-L\left(\chi^{\nu_2\theta_5}\partial_{\theta_5}\phi\right)n^{\mu_2} \chi^{\mu_1\nu_1}-L\left(\chi^{\mu_1\theta_3}\partial_{\theta_3}\phi\right)\left(\chi^{\nu_1\theta_4}\partial_{\theta_4}L\right)n^{\mu_2}n^{\nu_2}\\
&-L\left(\chi^{\mu_1\theta_3}\partial_{\theta_3}L\right)\left(\chi^{\nu_1\theta_4}\partial_{\theta_4}\phi\right)n^{\mu_2}n^{\nu_2}+L^2\left(\chi^{\mu_1\theta_3}\partial_{\theta_3}\phi\right)\left(\chi^{\nu_1\theta_4}\partial_{\theta_4}\phi\right)n^{\mu_2}n^{\nu_2}\bigg]\\
&+n^{\sigma_4}\Gamma_{\sigma_4,\mu_1\nu_1}\bigg(-2\partial_{\mu_2}\phi~\partial_{\nu_2}L-2\partial_{\mu_2}L~\partial_{\nu_2}\phi-2L~\partial_{\mu_2}\partial_{\nu_2}\phi+2\partial_{\mu_2}\phi~\partial_{\nu_2}\phi\\
&+L\chi^{\alpha_2\theta_7}\partial_{\theta_7}\phi~\Gamma_{\alpha_2,\mu_2\nu_2}\bigg)\bigg[-\left(\chi^{\alpha\beta}\partial_{\alpha}L~\partial_{\beta}L\right)n^{\mu_2}n^{\nu_2}\chi^{\mu_1\nu_1}+\left(\chi^{\mu_2\theta_2}\partial_{\theta_2}L\right)n^{\nu_2} \chi^{\mu_1\nu_1}\\
&-\left(\chi^{\mu_1\theta_3}\partial_{\theta_3}L\right)n^{\nu_2} \chi^{\mu_2\nu_1}-\left(\chi^{\nu_1\theta_4}\partial_{\theta_4}L\right)n^{\mu_2} \chi^{\mu_1\nu_2}+\left(\chi^{\nu_2\theta_5}\partial_{\theta_5}L\right)n^{\mu_2} \chi^{\mu_1\nu_1}\\
&+\left(\chi^{\mu_1\theta_3}\partial_{\theta_3}L\right)\left(\chi^{\nu_1\theta_4}\partial_{\theta_4}L\right)n^{\mu_2}n^{\nu_2}+\chi^{\mu_1\nu_2}\chi^{\mu_2\nu_1}-\chi^{\mu_1\nu_1}\chi^{\mu_2\nu_2}\\
&+L\left(\chi^{\alpha\beta}\partial_{\alpha}\phi~\partial_{\beta}L\right)n^{\mu_2}n^{\nu_2}\chi^{\mu_1\nu_1}-L^2\left(\chi^{\alpha\beta}\partial_{\alpha}\phi~\partial_{\beta}\phi\right)n^{\mu_2}n^{\nu_2}\chi^{\mu_1\nu_1}\\
&+L\left(\chi^{\alpha\beta}\partial_{\alpha}L~\partial_{\beta}\phi\right)n^{\mu_2}n^{\nu_2}\chi^{\mu_1\nu_1}-L\left(\chi^{\mu_2\theta_2}\partial_{\theta_2}\phi\right)n^{\nu_2} \chi^{\mu_1\nu_1}+L\left(\chi^{\mu_1\theta_3}\partial_{\theta_3}\phi\right)n^{\nu_2} \chi^{\mu_2\nu_1}\\
&+L\left(\chi^{\nu_1\theta_4}\partial_{\theta_4}\phi\right)n^{\mu_2} \chi^{\mu_1\nu_2}-L\left(\chi^{\nu_2\theta_5}\partial_{\theta_5}\phi\right)n^{\mu_2} \chi^{\mu_1\nu_1}\\
&-L\left(\chi^{\mu_1\theta_3}\partial_{\theta_3}\phi\right)\left(\chi^{\nu_1\theta_4}\partial_{\theta_4}L\right)n^{\mu_2}n^{\nu_2}-L\left(\chi^{\mu_1\theta_3}\partial_{\theta_3}L\right)\left(\chi^{\nu_1\theta_4}\partial_{\theta_4}\phi\right)n^{\mu_2}n^{\nu_2}\\
&+L^2\left(\chi^{\mu_1\theta_3}\partial_{\theta_3}\phi\right)\left(\chi^{\nu_1\theta_4}\partial_{\theta_4}\phi\right)n^{\mu_2}n^{\nu_2}\bigg]\\
\end{split}
\end{equation}
\normalsize

\bibliography{FluidEntropyFinal.bib}
\bibliographystyle{JHEP}

\end{document}